\documentclass[twocolumn]{amsart}

\usepackage{amsmath,amssymb}
\usepackage{graphicx}

\def\ee{\mathrm e}

\def\cB{\mathcal {W}}
\def\cW{\mathcal {W}}

\def\Rconf{\mathcal {M}}

\def\Vol{{p}}

\def\cM{\mathcal {M}}
\def\cR{\mathcal {R}}
\def\cMi#1{\mathcal {M}^{(#1)}}

\def\ZZ{{\mathbb Z}}

\def\RR{{\mathbb R}}

\def\cE{{\mathcal{E}}}
\def\hcE{{\hat{\mathcal{E}}}}

\def\agg{{\mathrm{agg}}}

\def\book#1{\rm{#1}, }
\def\paper#1{\textit{#1}, }
\def\jour#1{\rm{#1}, }
\def\yr#1{({\rm{#1}) }}
\def\vol#1{\textbf{#1}}
\def\pages#1{\rm{#1}}

\def\by#1{{\rm{#1}, }}

\begin{document}

\twocolumn[
\title{Measuring Agglomeration of Agglomerated Particles Pictures}
%
%%%%% Author(s)
% List all authors in the order you would like them to appear.
% Syntax: \author{author label}{author name}{author email address}
% The author labels must all be distinct and consist of alphabetic characters only.
% Author names: first name in full, other initial(s), family name in full.
% Email addresses: use (at) instead of @.
%\author{SM}{Shigeki Matsutani}{matsutani.shigeki(at)canon.co.jp}
%\author{YS}{Yoshiyuki Shimosako}{shimosako.yoshiyuki(at)canon.co.jp}
%=================================================
\author{Shigeki Matsutani, Yoshiyuki Shimosako}
\date{June 6, 2013}
%=================================================
% If the list of author names is too long for a running head, use the following:
% \authorabbrev{John Doe et al.}
%
%%%%% Affiliation(s)
% Syntax: \affiliation{list of labels of authors}{address}
% The labels must be comma separated without space.
%\affiliation{SM,YS}{Simulation \& Analysis R\&D Center,
%Canon Inc., 3-30-2, Shimomaruko Ohta-ku, Tokyo, Japan}

%%%%% Abstract
\begin{abstract}%
%\abstract{%
 In this article, we introduce a novel 
geometrical index $\delta_\agg$, which is
 associated with the Euler number and is obtained by an 
image processing procedure
for a given digital picture of
aggregated particles   
such  that $\delta_\agg$
exhibits the degree of the agglomerations of the particles.
In the previous work (Matsutani, Shimosako, Wang, Appl.Math.Modeling
{\bf{37}} (2013), 4007-4022), 
we proposed an algorithm 
to construct a picture of
agglomerated particles as a Monte-Carlo simulation whose agglomeration
degree is controlled by $\gamma_\agg \in (0,1)$. 
By applying the image processing procedure to the
pictures of the agglomeration particles 
constructed following the algorithm,
we show that $\delta_\agg$ statistically
reproduces the agglomeration parameter $\gamma_\agg$.
%}
%\\
%\keywords{%
agglomeration, digital image processing procedure,
Euler number
%}
\end{abstract}%

%\begin{document}
\maketitle
]
%\keywords{%
%}
%%%%% author-defined commands and environments
% \newcommand{\F}{\mathbb{F}}
% \theoremstyle{ap-thm} %% italic
% \theoremstyle{ap-rem} %% upright
% \newtheorem{problem}{Problem}

\section{Introduction} 
\label{sec:Int}
Nano-composite materials have a promising future from industrial viewpoints,
since in the materials, geometrical properties in micro-scale
 play crucial roles and
generate novel and various macro-material properties.
By controlling the geometrical properties or shapes,
we can design the macro-material properties drastically.
Following Kelvin's philosophy of science 
 \cite{K}\footnote{
Kelvin wrote his philosophy of science,
``In physical science the first essential step in the direction of
learning any
subject is to find principles of numerical reckoning and practicable methods
for measuring some quality connected with it. I often say that when you can
measure what you are speaking about, and express it in numbers, you know
something about it; but when you cannot measure it, when you cannot express
it in numbers, your knowledge is of a meagre and unsatisfactory kind; it may
be the beginning of knowledge, but you have scarcely in your thoughts
advanced to the state of Science, whatever the matter may be.''\cite{K}},
it is quite important
to evaluate such geometrical properties or shapes 
if one needs to control them using the scientific knowledge.

On the other hand, the smaller the particle is, the larger the effect
of the surface energy is. It means that small particles
are apt to aggregate or agglomerate in general because
the agglomeration and the aggregation of the particles
decrease the total surface energy and contribute to the
stability of the system.
When we handle materials consisting of nano-particles, 
the agglomeration and the aggregation are ones of the  
most important shapes since they sometimes have an effect on
 the generations of the macro-materials properties.
In this article, we focus on them.
It is said that the aggregation is due to chemical effects whereas
the agglomeration comes from physical effects.
Since in a computational model, there is no difference between them,
we call both agglomeration in this article, though
in spatial point analysis \cite{IPSS}, 
the aggregation is chosen in general.

In the article \cite{MSW},
in order to find the agglomeration effect in
the electric conductivity of the nano-composite material,
we study the electric conductivity in an agglomerated
continuum percolation model and show that the 
agglomeration of particles affects the
macro-material properties.
The purpose of this article is to evaluate the agglomeration in
the binary digital images of agglomerated particles, e.g.,
of electron-microscopes.

For the same purpose,
so many evaluation methods and definitions of the agglomeration
are proposed to evaluate the agglomeration.
In spatial point analysis, the distribution of
the nearest distance particles, Clark-Evans index and so on are
considered \cite{CE,IPSS}.
%This problem is a quite old problem.
Further Miles considered the problem in \cite{Mi} and showed
the two-dimensional overlapping ratio of the random configurations 
in three dimensional space.
These investigations on the agglomeration have been done in
the framework of
the statistical analysis for a point pattern $\cR =\{p_i\in \RR^2\}$
which are given as statistical configurations of (finite) points.
In the analysis, 
$\cR_r
:= \overline{\bigcup_{p \in \cR} U_{r, p}}$ is investigated,
which is a configuration of disks whose centers are $\cR$,
where
$U_{\varepsilon, p}:=\{q \in \RR^2 \ |\ |q-p|<\varepsilon\ \}$.
The Euler number, the area and the perimeter of $\cR_r$
for several point processes $\cR$'s are computed as
morphological indices or the Minkowski characterization \cite{IPSS}.
When $\cR$ is given by the point process of the Poisson type, 
Stoyan, Kendall and Mecke studied their behaviors based
on the study of Miles \cite{Mi} and found that
\begin{equation}
   e(x) = (1-x) \ee^{-x}, \quad
   a(x) = \frac{1}{x} (1-\ee^{-x}), \quad
   \ell(x) = \ee^{-x}, \quad
\label{eq:Mink}
\end{equation}
where $e(x)$, $a(x)$, and $\ell(x)$ are the normalized versions of
the Euler number, the area and the perimeter of $\cR_r$,
and $x$ is a normalized radius $r$ \cite{SKM,MS}.
Mecke and Stoyan studied the difference among point patterns
given by different processes in terms of these behaviors \cite{MS}.
%Further by investigating the sequence of the Euler number of $\cR_r$,
%Mecke and Stoyan studied the geometrical (morphological) character
%of the point patterns or CPMs \cite{MS}.
Further Tscheschel and Stoyan also studied the statistical reconstruction
of random point patterns \cite{TS}.

However in the nano-composite materials consisting of nano-particles,
the particles themselves sometimes have complicated shapes, 
such as ellipsoids and rods,  as we investigated in \cite{MSW0}. 
In other words,
the configuration is not given by a point pattern with radius $r$
in general and further its parts are overlapped like (d) in
 Figures \ref{fig:2Dview0.2} and  \ref{fig:2Dview0.4}.
Hence it is basically an ill-posed problem to define the 
center points of actual agglomerated
particles in a given picture, e.g., of an electron-microscopes.

Thus it is natural to consider geometrical properties of the 
binary picture as a general geometrical object embedded in $\RR^2$. 

Recently MacPherson and Schweinhart proposed a novel method
which evaluates the complicatedness of  
the complicated geometric objects embedded in a
plane $\RR^2$ in terms of 
 the persistent homology  \cite{MPS}.
The persistent homology gives the homological quantities
of the persistent modules with real parameter \cite{EH,W}. 
It could be regarded as a generalization of
homotopical approach in traditional algebraic 
topology \cite{BT}, though the deformation does not preserve
homotopical properties. 
For a geometrical object $\cM \subset \RR^2$, we consider a 
family of objects with a real parameter $t \in [0,1]$, i.e., 
$\{\cM_t \ | t \in [0,1]\}$.

By considering union of the $\varepsilon$-neighborhood of each
point in $\cM$,
$\cM_\varepsilon := \overline{\bigcup_{p \in \cM} U_{\varepsilon, p}}$,
induced from the standard Euclidean topology,
MacPherson and Schweinhart evaluated the complexity of the geometrical
objects.
The persistent homology shows the distributions of topological changes 
generated by the persistent modules (vector spaces)
induced from $\cM_{t'} \subset \cM_t$ for $t' < t$.

As in \cite{MPS},
to investigate the effect from the standard topology of Euclidean
space and to evaluate the complexity, 
we use the one-parameter family of a deformed geometrical object,
and propose a digital image processing procedure which
characterizes the shapes in pictures of the electric microscope
in this article.
(In Section \ref{sec:EAC}, we give the list of assumed geometrical 
features of the pictures which we deal with.)
For an appropriate geometrical object $\cM$ in $\RR^2$
with a characteristic lengths $\ell_1$ and $\ell_2$, 
we also handle the family of geometrical objects 
$\{\cM_t= \overline{\bigcup_{p \in \cM} U_{t, p}} \ $ $
| t \in [\ell_1,\ell_2]\}$.
We define the cumulus of the absolute differential Euler number
(CADE) by,
\begin{equation}
\cE(\cM; \ell_2, \ell_1) := \int_{\ell_1}^{\ell_2} \left|
\frac{d\chi(\cM_t)}{dt}\right| dt,
\label{eq:Euler}
\end{equation}
where  $\chi(X)$ is the Euler number of $X$.
$\cE(\cM;\ell_2,\ell_1)$ evaluates how many topology changes occur for
the deformation $[\ell_1, \ell_2]$.

As we are concerned with the image processing procedure 
for images of the electron-microscopes,
we will customize $\cE(\cM;\ell_2,\ell_1)$ as $\hcE(\cM;\ell_2,\ell_1)$
for any binary pictures as an image processing procedure, which 
is shown in Section \ref{sec:EAC} more precisely.
Further in the nano-materials, there are several scales and one of them
is the size of the particles and 
the resolution of the digital picture is given by the pixel
size.  We fix $\ell_1$ and $\ell_2$ by
the pixel size and the
(average) radius $\rho$ of the particle respectively
to evaluate the agglomeration
and propose an agglomeration index, 
\begin{equation}
     \delta_\agg(\cM) := 
\frac{\alpha}{\hcE_{p(\cM)}}
(\hcE_{p(\cM)}- \hcE(\cM;\rho,a)),
\label{eq:delta}
\end{equation}
where $p(\cM)$ is the volume fraction of $\cM$ in the region $\cW$
$(\cM \subset \cW \subset \RR^2)$,
$\hcE_p$ is the average of a ``standard pattern of volume fraction 
$p$" as mentioned in Section \ref{sec:EAC}, 
and $\alpha$ is a normalized factor $1.2$.

In order to estimate our agglomeration parameter $\delta_\agg$,
we performed Monte-Carlo simulations for the binary agglomeration
configurations of particles whose degree of the agglomeration
 is parameterized by $\gamma_\agg$,
since in the article \cite{MSW},
we proposed a statistical model
which numerically generates the agglomeration of particles
controlled by the  parameter $\gamma_\agg$
in order to investigate
the properties of the agglomerated continuum percolation models.
In Section \ref{sec:AC}, 
we review the algorithm following the article \cite{MSW}.
For a given parameter $\gamma_\agg \in [0, 1]$,
we can statistically construct the infinitely many configurations with 
the same level of the agglomerations.
As the continuum percolation model is the same as a germ-grain model
in the study for the point process \cite{IPSS},
there are several other algorithms to construct aggregational
germ-grain models, such as Neyman-Scott processes \cite{IPSS},
though they are different from ours; 
%These methods are based on an algorithm using
% artificial clusters, whereas 
ours is for the actual pictures of electron-microscopes of
the agglomerated nano-composite materials with the radius 
$\rho$ as mentioned in \cite{MSW}.
We apply the index $\delta_\agg$ to evaluate the agglomeration of 
the agglomerated configuration
which is generated by the forward method in \cite{MSW}.
Then the relevancy between 
$\delta_\agg$ and $\gamma_\agg$ is shown in Section \ref{sec:NCR}, i.e., in
Figure \ref{fig:deltagamma} and Table \ref{tbl:deltavsgamma}.
We also mention the relation between our $\delta_\agg$ and 
the well-established Clark-Evans index in Section \ref{sec:NCR}.

\section{Agglomerate configuration}
\label{sec:AC}

\begin{figure}[htbp]
 \begin{tabular}{cc}
  \begin{minipage}[t]{0.45\hsize}
   \begin{center}
    \includegraphics[height=\hsize, clip]{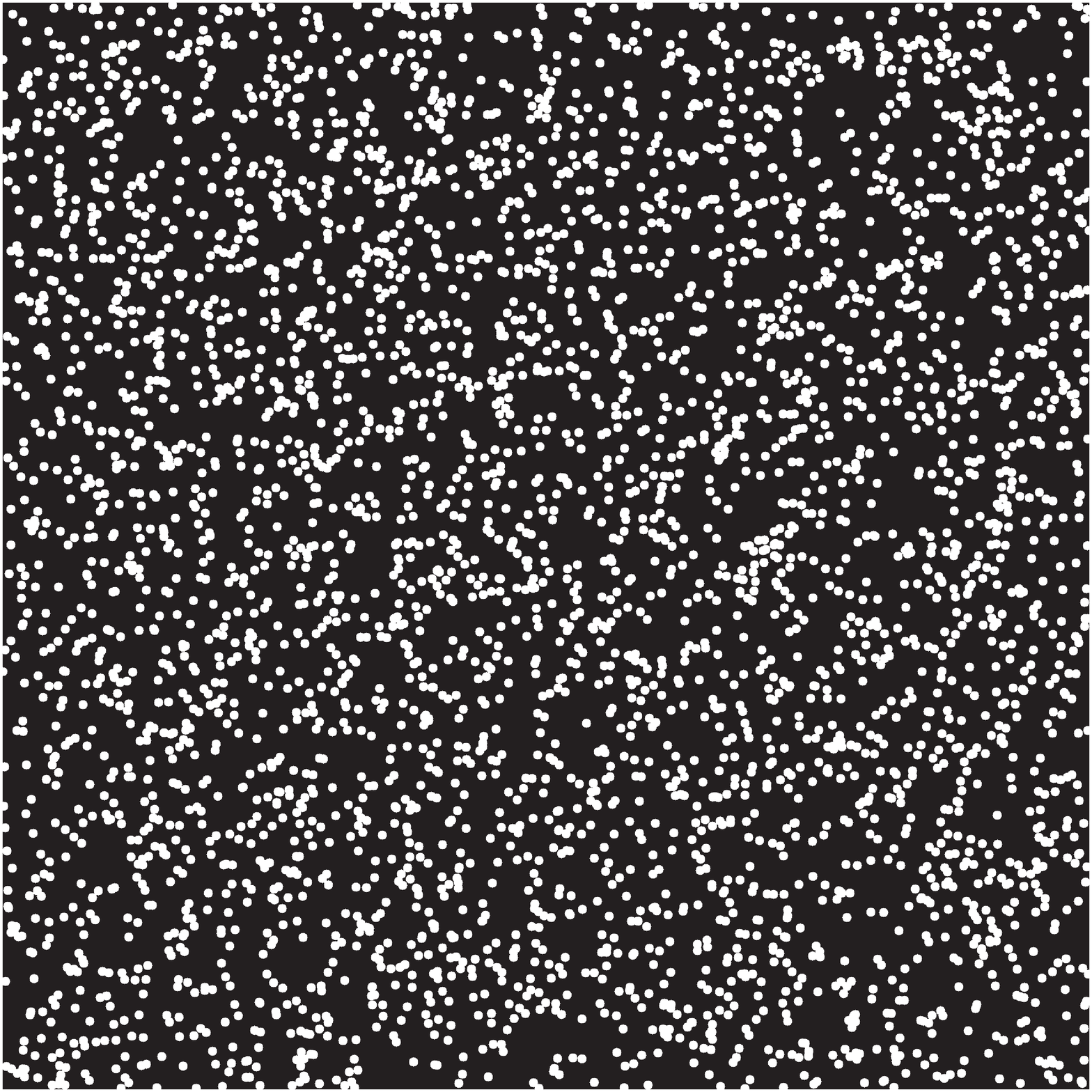}
    (a)
   \end{center}
   \begin{center}
    \includegraphics[height=\hsize, clip]{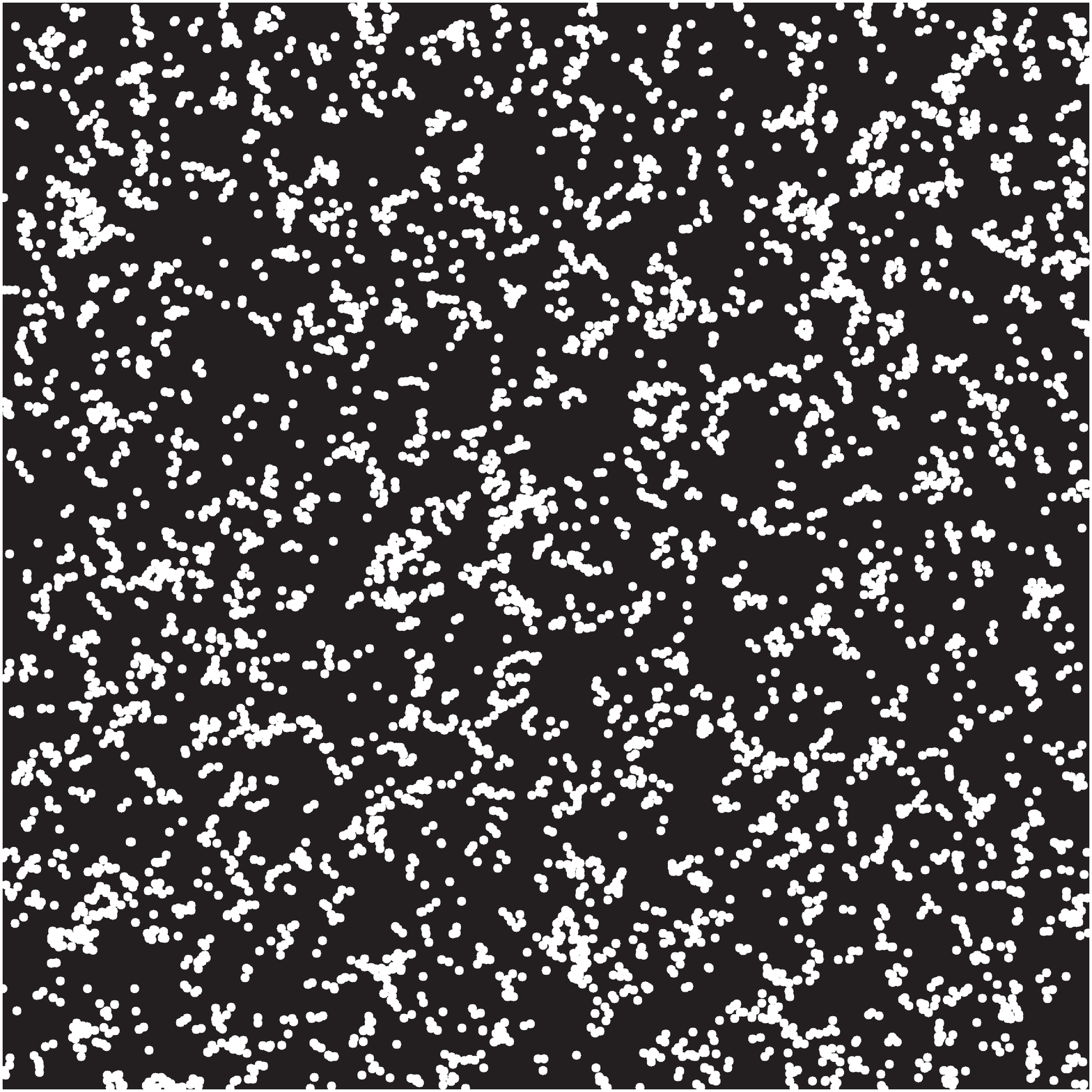}
    (c)
   \end{center}
  \end{minipage} 
  \begin{minipage}[t]{0.45\hsize}
   \begin{center}
    \includegraphics[height=\hsize, clip]{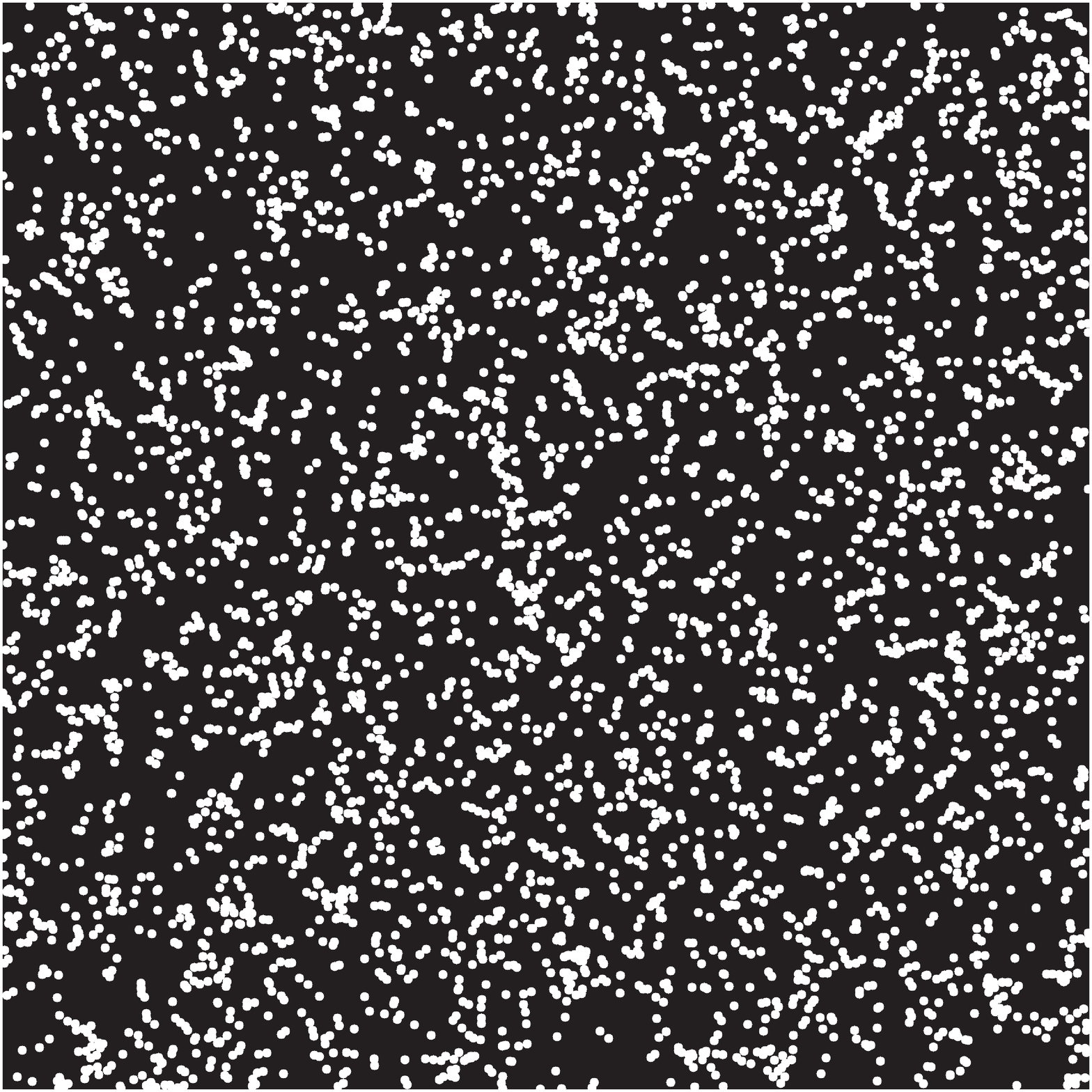}
    (b)
   \end{center}
   \begin{center}
    \includegraphics[height=\hsize, clip]{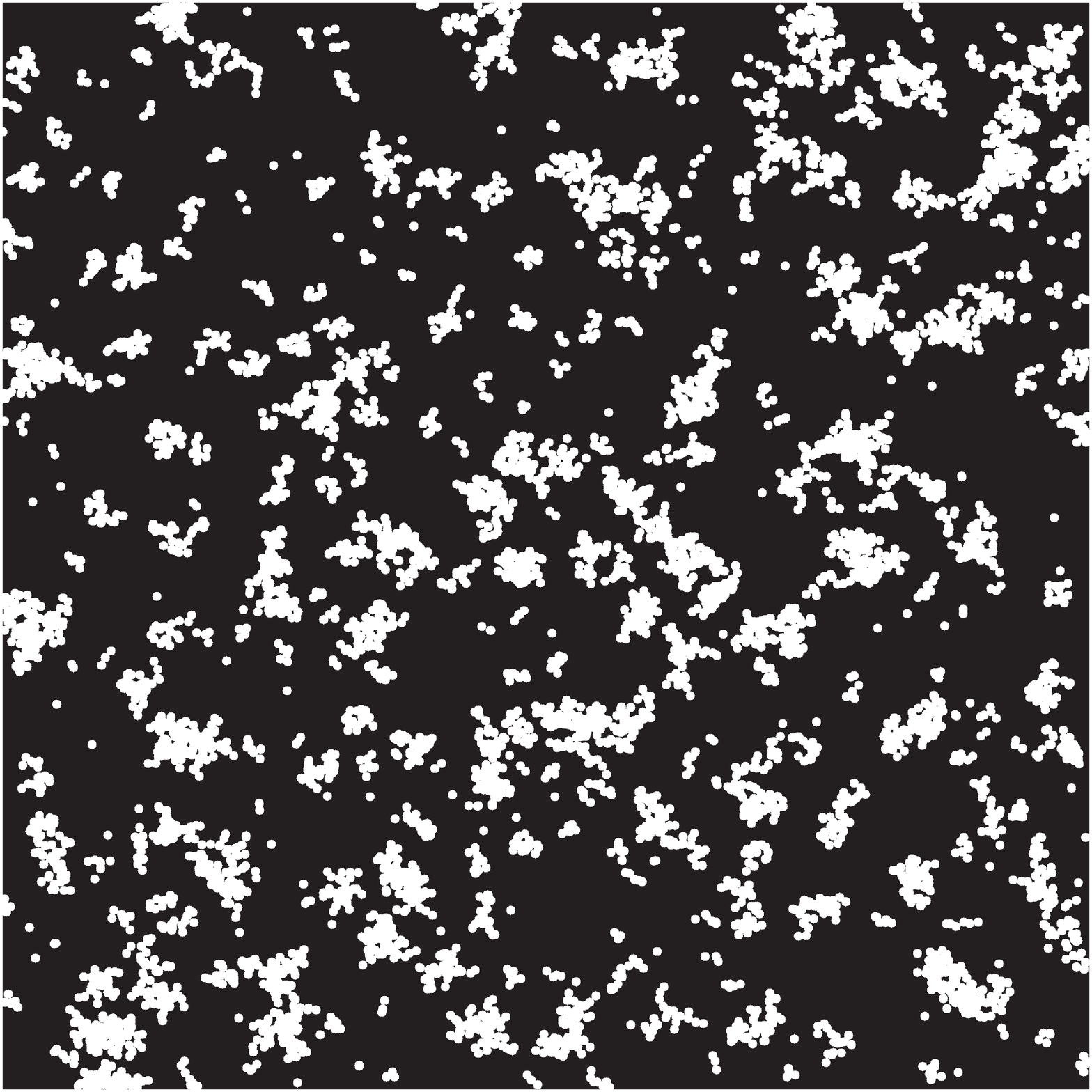}
    (d)
   \end{center}
  \end{minipage} 
 \end{tabular}
\caption{\label{fig:2Dview0.2} 
The agglomerated configurations of $p=0.2$:
These (a), (b), (c), and (d) 
show the configurations with the agglomeration parameter 
$\gamma_\agg$ $=0.0, 0.3, 0.6$ and $0.9$  respectively.
}
\end{figure}

\begin{figure}[htbp]
 \begin{tabular}{cc}
  \begin{minipage}[t]{0.45\hsize}
   \begin{center}
    \includegraphics[height=\hsize, clip, angle=270]{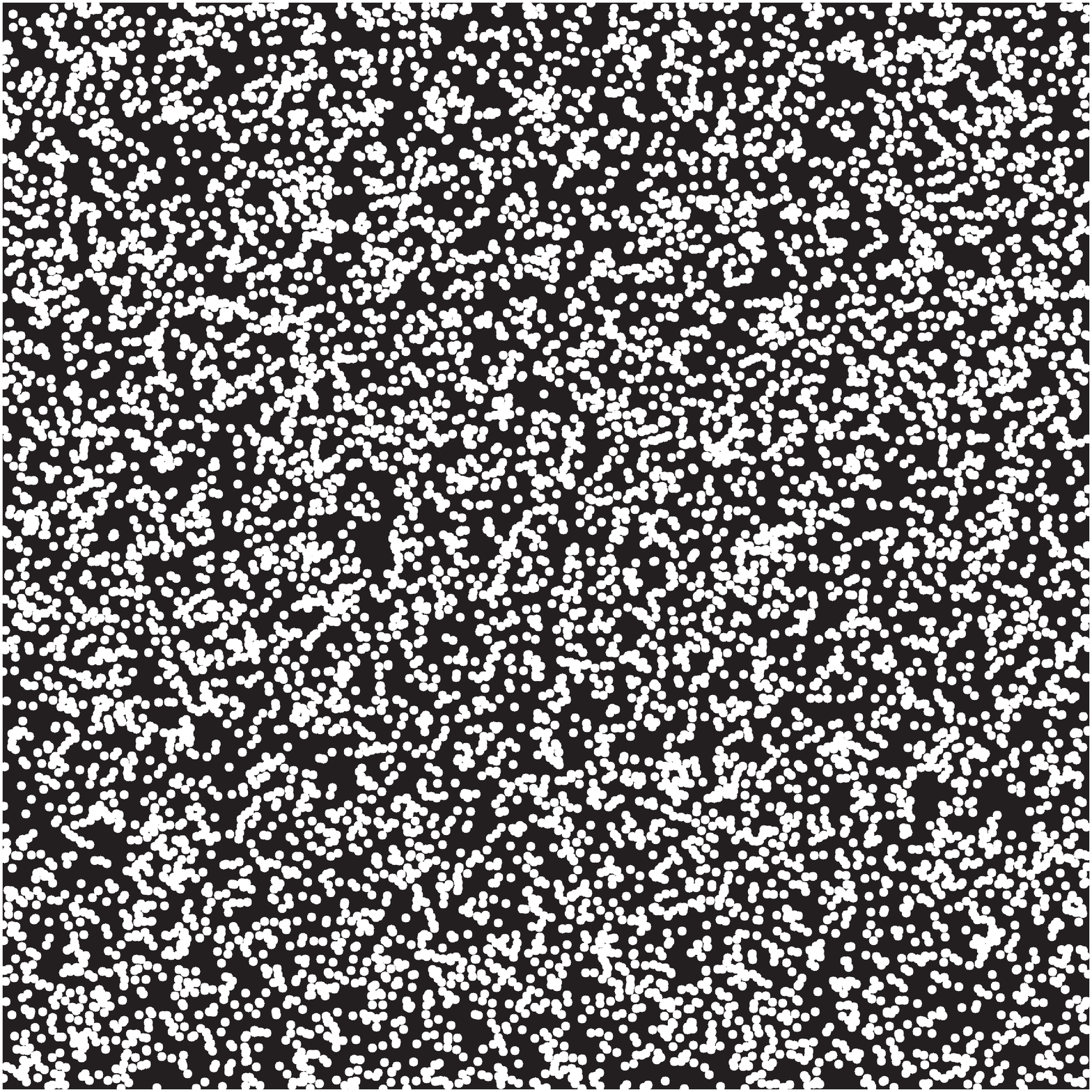}
    (a)
   \end{center}
   \begin{center}
    \includegraphics[height=\hsize, clip, angle=270]{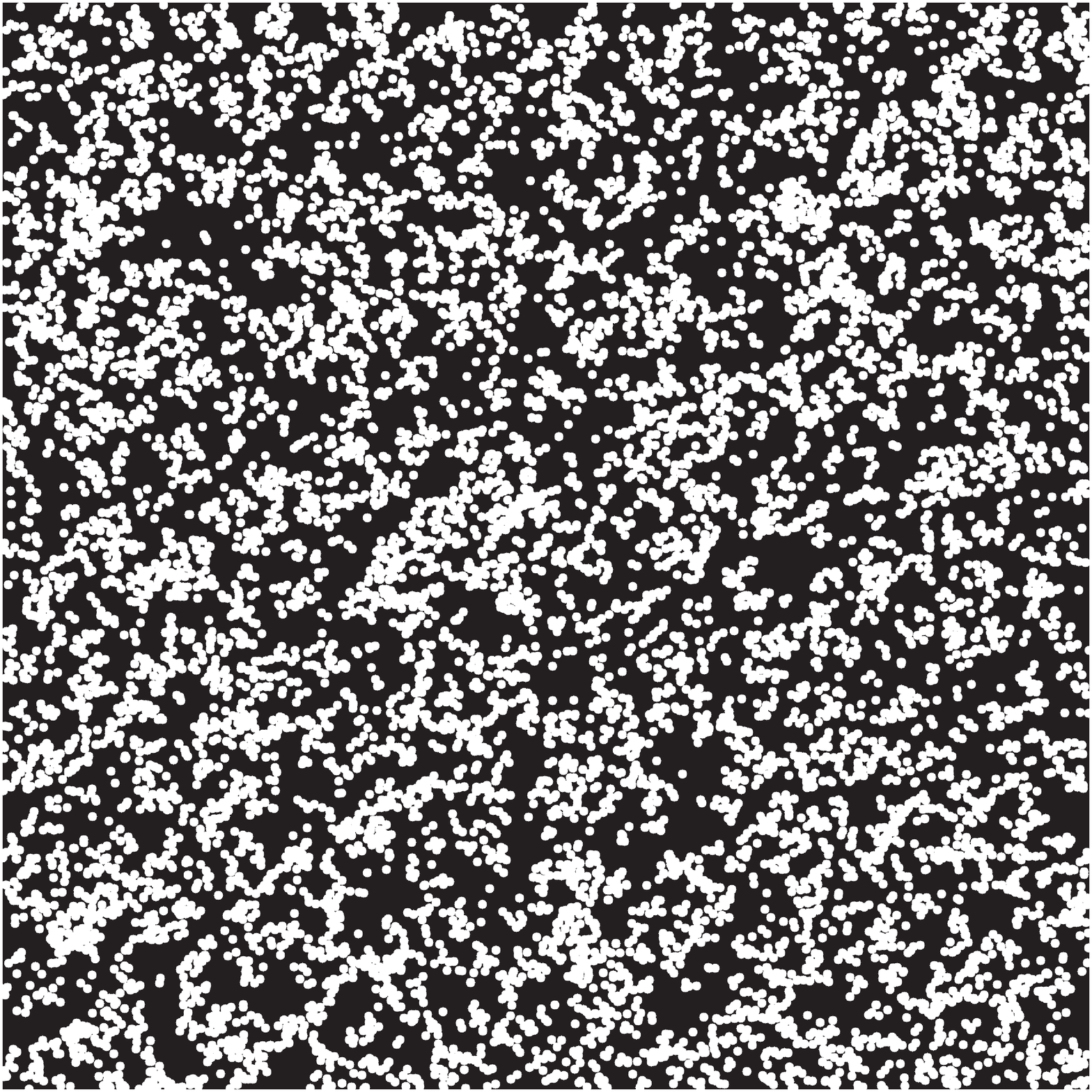}
    (c)
   \end{center}
  \end{minipage} 
  \begin{minipage}[t]{0.45\hsize}
   \begin{center}
    \includegraphics[height=\hsize, clip, angle=270]{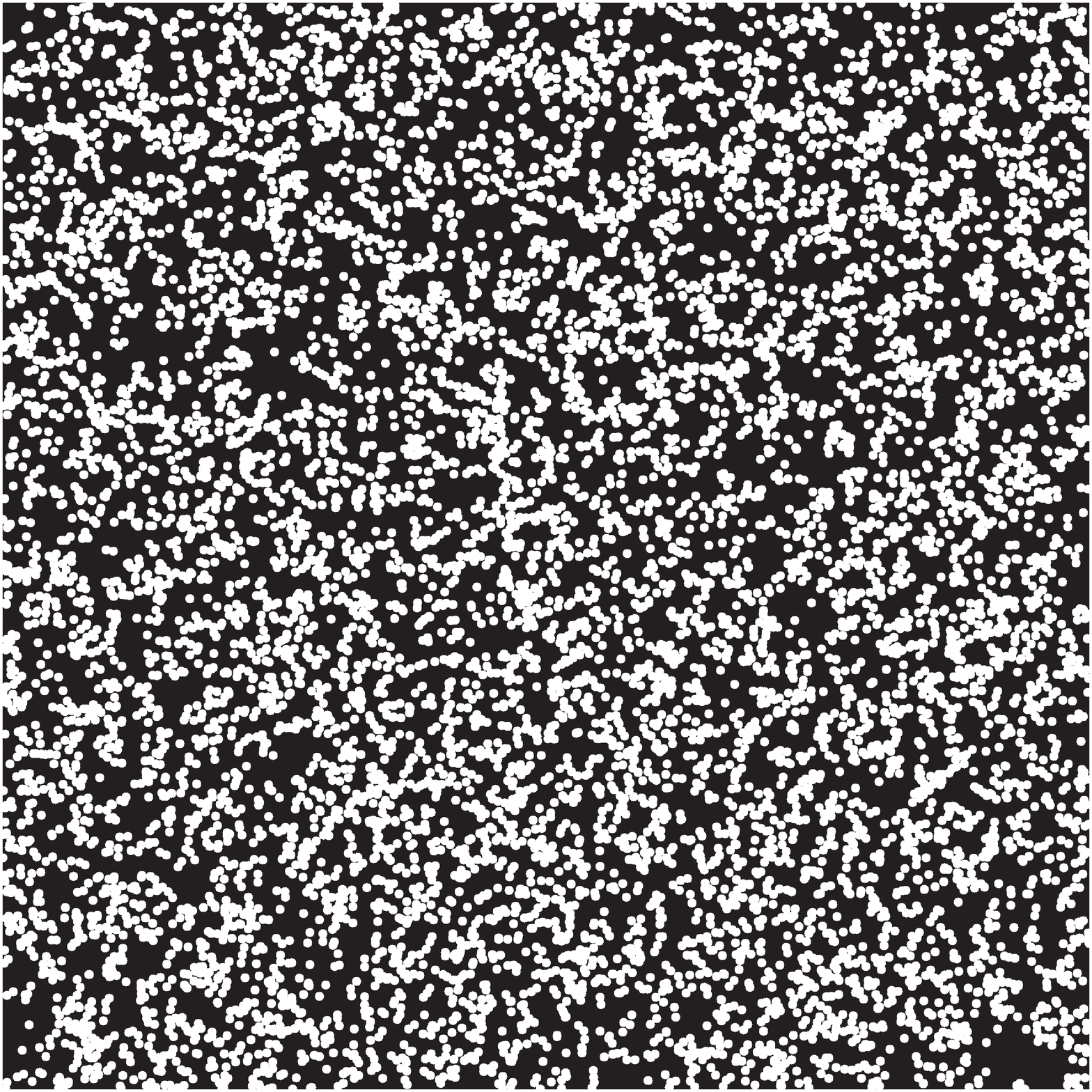}
    (b)
   \end{center}
   \begin{center}
    \includegraphics[height=\hsize, clip, angle=270]{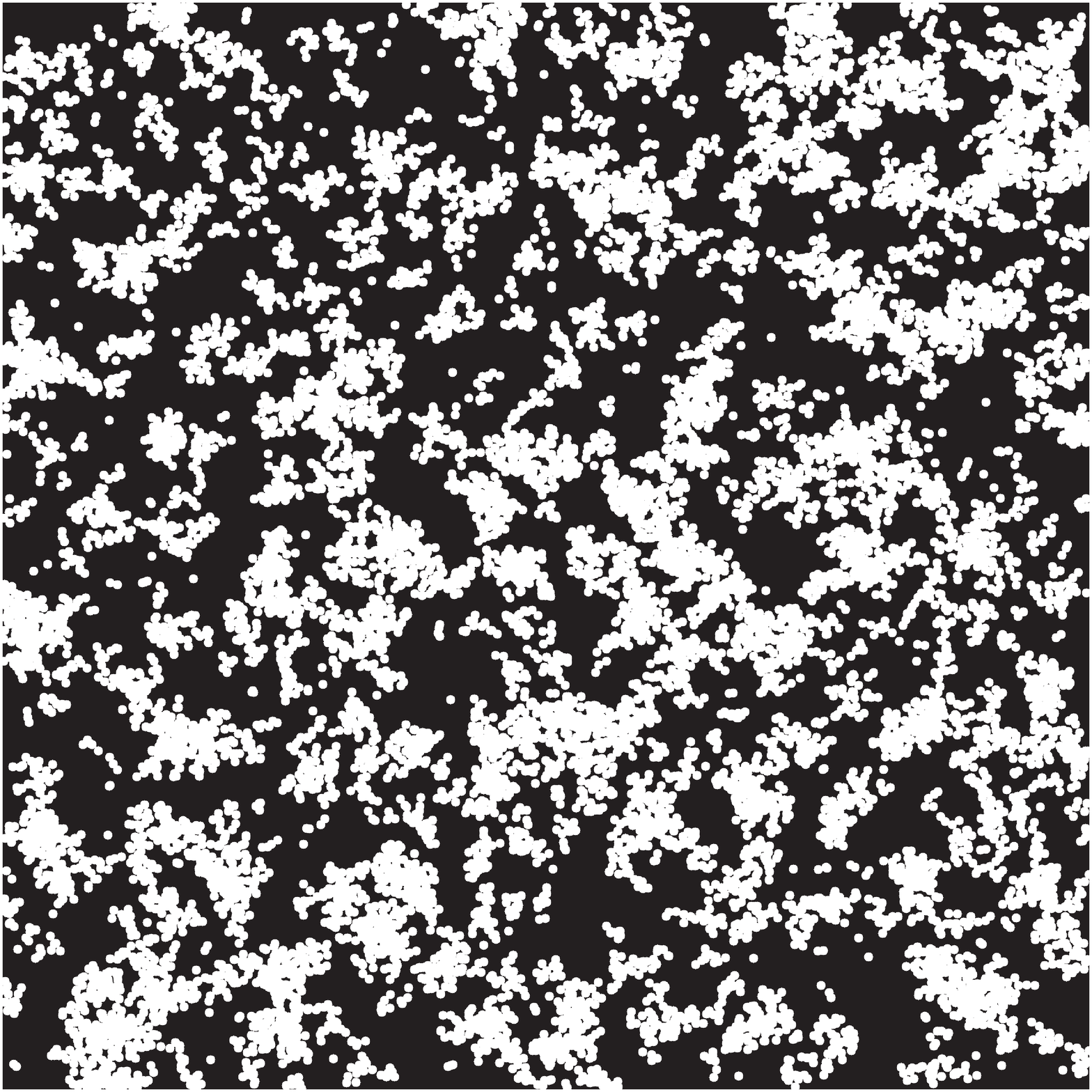}
    (d)
   \end{center}
  \end{minipage} 
 \end{tabular}
\caption{\label{fig:2Dview0.4} 
The agglomerated configurations of $p=0.4$:
These (a), (b), (c), and (d) 
show the configurations with the agglomeration parameter 
$\gamma_\agg$ $=0.0, 0.3, 0.6$ and $0.9$  respectively.
}
\end{figure}

In order to explain what is the agglomeration that we are concerned with,
we show the agglomeration configurations in computer science,
which we handled in \cite{MSW}.
In the article \cite{MSW}, we proposed a construction of the agglomerated 
continuum percolation models which apparently recover geometric
properties of real nano-particles, though there are several
other agglomerated percolation models such as Neyman-Scott processes 
\cite{IPSS,SKM,TS}.
Since our method has a single parameter $\gamma_\agg$ besides
a typical length $\rho$
whereas others are given as point processes with several parameters, 
we believe that ours is a natural model for the 
actual pictures of electron-microscopes of
the agglomerated nano-composite materials with the radius $\rho$.
As shown in Figures \ref{fig:2Dview0.2} and \ref{fig:2Dview0.4}, 
we have the agglomerated configuration of particles
depending upon a agglomeration parameter $\gamma_\agg\in [0,1]$.
In this section, we show the geometrical setting of agglomerated
continuum percolation model in \cite{MSW}, which is  modeled by
the agglomerated clusters in nature.

We set particles parameterized by their center positions $(x,y)$ into
a box-region $\cB:=[0,L]\times [0,L]$ at random and
get a configuration $\Rconf_n$ as a model of continuum percolation.
The particle corresponds to a disk with the same radius $\rho$,
$B_{x_i, y_i}:= \{ (x, y) \in \cB \ | \ |(x,y)-(x_i, y_i)| \le \rho\}$.
The configuration $\Rconf_n$ is given by 
$\Rconf_n:=\bigcup_{i=1}^n B_{x_i, y_i}$.

\begin{figure}[htbp]
\includegraphics[scale=0.35]{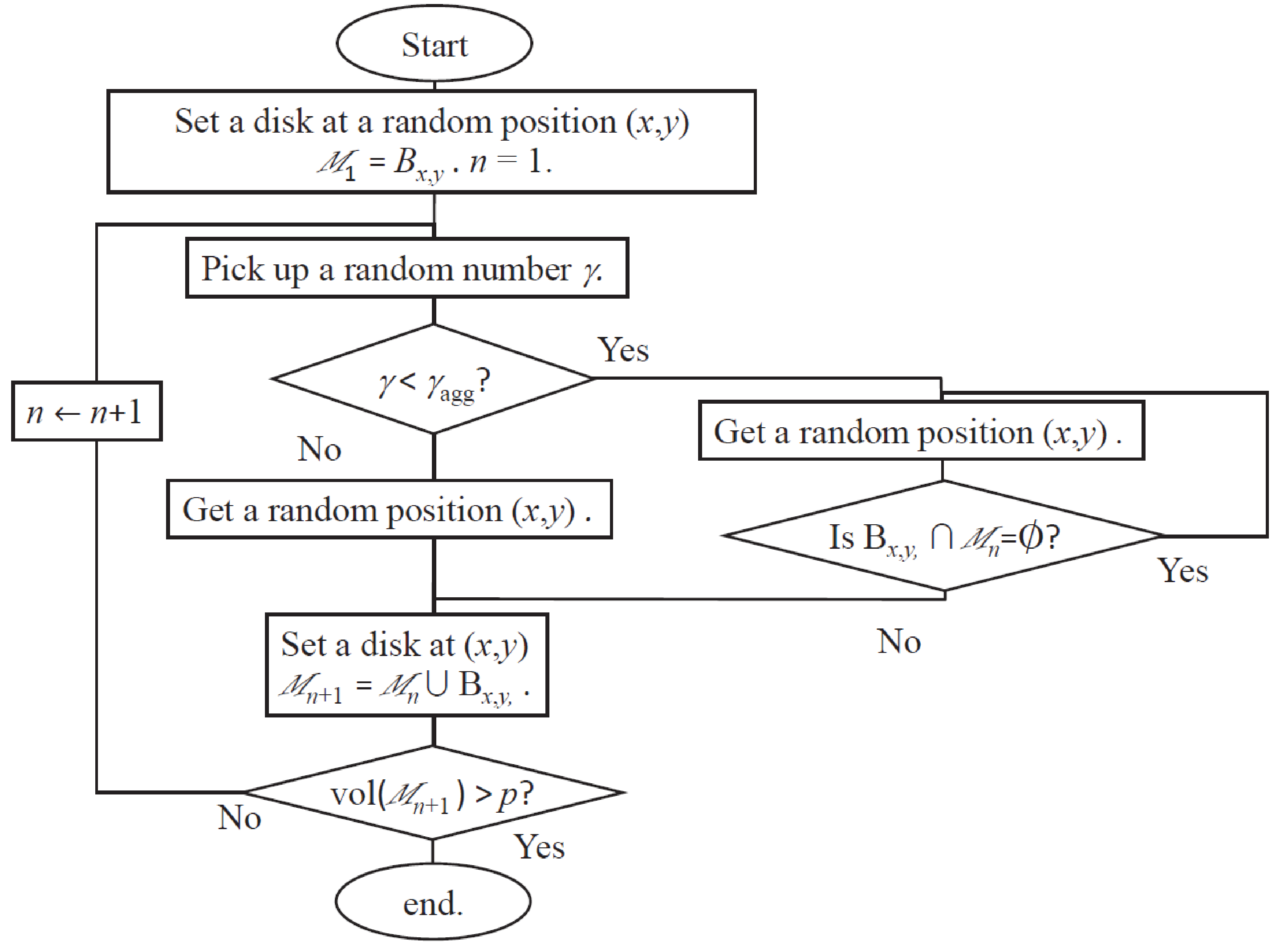}
\caption{\label{fig:flowchart} The flowchart of 
the algorithm which constructs the agglomerated configurations.}
\end{figure}
The flowchart in Figure \ref{fig:flowchart} illustrates the algorithm.
As the initial state, the configuration $\Rconf_0$ has no particle.
As the first step, for a uniform random position $(x,y) \in \cB$,
we set a particle $B_{x,y}$ whose center is $(x,y)$ and
the radius is $\rho$, i.e., $\Rconf_1:=B_{x,y}$.
%$:=\{(x',y') \in \cB\ | \ |(x',y')-(x,y)|\le \rho\}$.

For the $(n+1)$-th step, we take a position $(x,y)$ at uniform random
in $\cB$, and another random parameter $\gamma$ at uniform random
in $[0,1]$.  If $\gamma$ is greater than $\gamma_\agg$,
we set $\Rconf_{n+1}:=\Rconf_{n} \bigcup B_{x,y}$.
We now allow the particles to overlap each other.

For the case $\gamma \le \gamma_\agg$,
we first check whether the disk $B_{x,y}$
is connected with the previous configuration $\Rconf_n$  or not. 
For the case $\Rconf_n\bigcap B_{x,y}\neq \emptyset$, we employ the
position and set $\Rconf_{n+1}:=\Rconf_n \bigcup B_{x,y}$.
Otherwise or $\Rconf_n\bigcap B_{x,y}= \emptyset$, 
we abandon the position and go on to take another
uniformly random position $(x,y)$ in $\cB$ until we find 
the position which supplies a connected particle $B_{x,y}$ with $\Rconf_n$.

In other words, for the case $\gamma \le\gamma_\agg$,
the added particle must be connected with the previous configuration 
$\Rconf_n$. Thus, $\gamma_\agg$ stands for the agglomeration of the 
particle system. 

By monitoring the total volume fraction which is
a function of $\Rconf_n$ and is denoted by $\Vol(\Rconf_n)$, 
we go on to put the particles as long as
$\Vol(\Rconf_n) \le p$ for a given volume fraction $p$.
We find the step $n(p)$ such that
$\Vol(\Rconf_{n(p)-1}) \le p$ and
$\Vol(\Rconf_{n(p)}) > p$.
Since we assume that the difference between $\Vol(\Rconf_{n(p)-1})$ 
and $\Vol(\Rconf_{n(p)})$ is  sufficiently small, 
we regard $\Vol(\Rconf_{n(p)})$ as the volume fraction $p$ itself
hereafter under this accuracy.

Since in the Monte-Carlo method,
we use the pseudo-randomness to simulate the random
configuration $\Rconf_{n(p)}$ for given $p$ and $\gamma_\agg$,
the configuration $\Rconf_{n(p)}$ depends upon the
seed $i_S$ of the pseudo-randomness which we choose.
We let it be denoted by $\Rconf_{\gamma_\agg, p, i_S}$ or
its statistical quantity by $\Rconf_{\gamma_\agg, p}$.

For sufficiently large $L$ and $L'$ ($\rho \ll L' < L$)
 and for a window
$\cB_{(x,y)}' := [x,x+L']\times[y, y+L']\subset \cB$,
the volume fraction $p'(x,y)$ in $\cB_{(x,y)}'$
is proportional to 
the number of particles $N'(x,y)$ in $\cB_{(x,y)}'$, i.e.,
there is a constant number $\lambda$ such that
$p'(x,y) = \lambda N'(x,y)$.
Further 
$\Rconf_{\gamma_\agg, p, i_S}$ is isotropic and independently
scattered and thus the asymptotic behavior of
$\Rconf_{\gamma_\agg, p, i_S}$ is a kind of
Poisson process \cite[p.66]{IPSS} though we have the condition
$L'\gg \rho$.

Further though our algorithm for $\gamma_\agg >0$
is not Markov process because each
step depends on the previous configuration, we should note
 that it preserves
a hierarchical structure for fixed $i_S$ and $p>p'$,
$$
\Rconf_{\gamma_\agg, p', i_S} \subset \Rconf_{\gamma_\agg, p, i_S}.
$$

As we will show the assumed geometric properties of pictures in
Section \ref{sec:EAC}, the pictures which we will deal with are illustrated
in Figures \ref{fig:2Dview0.2}
and  \ref{fig:2Dview0.4}, which are obtained 
following our algorithm.
Our more concrete aim of this article is to recover the parameter
$\gamma_\agg$ for a given configuration $\Rconf_{\gamma_\agg, p}$
in a statistical meaning.
More precisely, our study is to find  statistical
monotone functions of $\gamma_\agg$ and to show that one of them
is $\delta_\agg$ in (\ref{eq:delta}).

\section{Evaluation of Agglomeration for 
a configuration $\Rconf_{p,\gamma_\agg} \subset \cW$}
\label{sec:EAC}

%\subsection{Evaluation of Agglomeration}

As we are concerned with the evaluation method as 
a digital image processing
procedure \cite{P}, in this section,
we illustrate our algorithm for a picture which only has binary values.
It is natural that we assume
the configuration $\cM$ (implicitly $\Rconf_{p,\gamma_\agg, i_S}$
and a picture of nano-composite material in an electron-microscope) 
has the following structures:

\begin{enumerate}

\item
$L$ is sufficiently larger than $\rho$ so that
the particles of $\cM$ are a representative
of sufficiently randomized configurations; 
we could assume the Euclidean invariance (translation, rotation and
inversion) statistically;
after averaging them, the physical and geometrical quantities
are invariant for any Euclidean action $\mathrm{E}(2)$ up to
the statistical deviation.
If the deviation is not small, 
we could consider the series of $\{\Rconf_{p,\gamma_\agg, i_S} \ | \
i_S \}$. (It means that for the case of the
pictures of the electron-microscopes,
we could assume that the researchers prepare the
series of pictures of a material or materials which are produced in the same
conditions.)

\item
It is assumed that the volume fraction is less than 
the percolation threshold of two dimensional continuum percolation
models.
(For the case of nano-composite material which
is based upon the percolation theory, 
the volume fraction around the percolation threshold
$0.2 \sim 0.3$ in three dimensional percolation models
is concerned, which is far less than the percolation
 threshold of two dimensional case $0.5 \sim 0.7$.)

\item
There are three sizes of the system or the picture 
$\Rconf_{p,\gamma_\agg}$ (and the digital image of 
nano-composite material of an electron-microscope);

\begin{enumerate}
\item the (average) size of particles, which is given by
$\rho$,

\item the analyzed size of the system, which is, now, given by $L$
as mentioned above, and

\item the pixel size $a$, which 
is also controlled so that we can discriminate the particles in
concerned resolution.

\end{enumerate}
\end{enumerate}

Under these assumptions, we consider  geometry of $\cM$.
It is known that the $\varepsilon$-neighborhood, 
$\cM_\varepsilon = \overline{\bigcup_{p \in \cM} U_{\varepsilon, p}}
\bigcap \cB$,
can be realized by
the so-called level set method in computer science \cite{S}.
Let $d: \cB \to \RR$ be the signed distance from the boundary
$\partial \cM$ so that the outer side is assigned to the positive distance
and the inner side is to the negative one,
and then the geometrical object in the level set method
can be regarded as $\mathcal{L}_t = d^{-1}(t)$. 
$\cM_t$ of $(t>0)$ is equal to
$d^{-1}([0,t])\bigcup\cM$ and 
$\mathcal{L}_t = \partial \cM_t$.
For $t<0$ case, 
$\mathcal{L}_t = \partial 
(\overline{\bigcup_{p \in \partial \cM} U_{t, p}})
\setminus \mathcal{L}_{-t}$.
Hence by means of the level set method, we can compute
the more precise geometrical properties beyond the pixel size resolution
even on the image defined over a subset of $\ZZ^2$.

However in the digital image processing procedure,
we investigate the geometrical object up to the pixel size resolution 
in general.
Further we must pay our attentions on the computational cost if
we apply our method to real problems in industry,
though level set function method requires higher computational
cost than a simple  digital image processing procedure.
Hence in this article, we use the thickening scheme in
the image processing procedure \cite{P} instead of the level set function.
Though the ordinary thickening scheme has anisotropic behavior,
it does not have a serious effect on the result because the configuration
itself is isotopic or rotational invariant. We use the modified
thickening scheme, which improves the
anisotropic behavior  shown in Section \ref{sec:NCR}.
Let $\cMi{i}$ be the $i$-th thickening of $\cM$ in $\cB$.
We modify the CADE (\ref{eq:Euler}) as an image processing procedure by
$$
\hcE(\cM;n_2a; n_1 a) :=   
\sum_{i=n_1 + 1}^{n_2} |\chi(\cMi{i})-\chi(\cMi{i-1})|.
$$
%For the case $na < \rho$, the anisotropy due to the thickening
%is not large and thus we regard  $\cMi{i}$ as $\cM_{ai}$.

In the persistent homology, the Betti number is handled in general.
Since the computational cost to the evaluation of
 the Euler number is not so high and
the Euler number could be compared with the results in \cite{IPSS,SKM},
we consider the behavior of the Euler numbers of 
$\cM_t$ in this article.
More precisely
though there is no guarantee that $\chi(\cM_{t})$ is equal to 
$\chi(\cMi{i-1})$ for $t \in [a(i-1/2), a(i+1/2))$,
we handle $\chi(\cMi{i-1})$;
as mentioned above,
in digital analysis,
we should basically neglect finer 
geometrical difference than the pixel size resolution and
we follow the principle.
In the complicated system, we believe that it is quite important
how many topology changes occur for the $i$-step,
and the difference of the Euler number can represent the behavior.

Further the agglomeration can be discriminated whether
the particles are connected or not.
From (\ref{eq:Mink}), if $L/\rho$ is  sufficiently large, 
even for $\gamma_\agg =0$ and small $p$,
$\hcE(\Rconf_{0,p}, n a,0)$ does not vanish $n>0$, 
in general, due to the randomness of the configurations. 
Further the behavior $\hcE(\cM;n_1a,n_2 a)$ of 
$n_1, n_2 \in [0, \rho/a)$ is quite important since
the agglomeration suppresses the topology change in the interval
as illustrated in 
 Figures \ref{fig:CADEvsTh} and \ref{fig:EulerPP}.

Due to the randomness of the configurations and the 
agglomeration, it is not so important
whether the Euler numbers increase or decrease, but 
 the topological change for the deformation is quite important.
We define the agglomeration parameter 
$\delta^{(n_1,n_2)}_{agg}$ in (\ref{eq:delta})
more precisely
\begin{equation}
     \delta^{(n_2,n_1)}_\agg(\cM) = 
\frac{\alpha}{\hcE^{(n_2,n_1)}_{p(\cM)}}
(\hcE^{(n_2,n_1)}_{p(\cM)}- \hcE(\cM; n_2a, n_1a)),
\end{equation}
where $p(\cM)$ is the volume fraction of $\cM$,
$\hcE^{(n_2,n_1)}_p$ is the average of the 
standard patterns of volume fraction $p$,
and $\alpha$ is a normalized factor $1.2$, which is chosen 
as a result of the comparison with 
$\gamma_\agg$ (see Table \ref{tbl:deltavsgamma}).
The standard pattern means the pattern of $\gamma_\agg=0$ 
with the same radius in the same window $\cB$.
Then $\delta^{(n_2,n_1)}_{\agg}(\Rconf_{\gamma_\agg,p})$
characterizes how many topological changes occur 
in the interval $(n_1 a, n_2 a)$ for the 
deformations for each particle in $\Rconf_{\gamma_\agg,p}$
by normalized by $\hcE^{(n_2,n_1)}_p$.

%Then we plotted $\delta_{\agg}(\Rconf_{i_S,\gamma_\agg,p})$ versus
%$\gamma_\agg$ in Figure \ref{fig:EulerPP}.
%Our evaluation $\delta_{\agg}(\Rconf_{i_S,\gamma_\agg,p})$ is relevant
%to $\gamma_\agg$. It means that our evaluate parameter recovers
%the agglomerated parameter $\gamma_\agg$.

\section{Numerical Computation and Results}
\label{sec:NCR}

Let us show the relevance between $\delta_p$ and $\gamma_p$
by the Monte-Carlo simulations following the algorithm mentioned in
Section \ref{sec:AC}.
Using the algorithm in Section \ref{sec:AC}, we have  ten 
pictures of agglomerated particles for each $\gamma_\agg = 0, 0.3, 0.6$
and $0.9$, and for each $p = 0.1, 0.2, 0.3$ and $0.4$
 by letting $L = 2400$ and $\rho = 10$ as in 
Figures \ref{fig:2Dview0.2}  and \ref{fig:2Dview0.4}.
%We computed the CADE by thickening of
%the digital image processing.

On the thickening to compute the CADE, we use two types
thickening process,
\begin{gather*}
\begin{array}{rl}
\mbox{type I: }& \ \Box \ \to \
\begin{matrix} 
     & \Box & \\
\Box & \Box & \Box \\
     & \Box & \\
\end{matrix},\\
\mbox{type II: }& \ \Box\ \to\
\begin{matrix} 
\Box & \Box & \Box \\
\Box & \Box & \Box \\
\Box & \Box & \Box \\
\end{matrix}, \\
\end{array}
\end{gather*}
such that we generate an octagon asymptotically
and approximates the area of the disks;
\begin{table}[htbp]
\begin{center}
\caption{The pattern of thickening}
\label{tbl:thickening}
\begin{tabular}{|c|c|c|r|r|}
\hline
steps &type & radius & area & n.of pixels  \\
\hline
1 &II & 0.5 & 0.785 & 1 \\
2 &I  & 1.5 & 7.065 & 9 \\
3 &I  & 2.5 & 19.625 & 21 \\
4 &I  & 3.5 & 38.465 & 37 \\
5 &II & 4.5 & 63.585 & 69 \\
6 &I  & 5.5 & 94.985 & 97 \\
7 &I  & 6.5 & 132.665 & 129 \\
8 &II & 7.5 & 176.625 & 185 \\
9 &I & 8.5 & 226.865 & 229 \\
10 &I & 9.5 & 283.385 & 277 \\
\hline
\end{tabular}
\end{center}
\end{table}
In other words, on the thickening process, we use the
deformation in digital process procedure for each point
which is given in Table \ref{tbl:thickening}.
Further for each point, we consider the thickening:
{\tiny{
\begin{gather*}
%             1234567890123456789012
\begin{array}{cccccccccccccccccccccc}
% 1 2 3 4 5 6 7 8 9 0 1 2 3 4 5 6 7 8 9 0
 & & & & & &0&0&0&0&0&0&0& & & & & & \\%1
 & & & & &0&9&9&9&9&9&9&9&0& & & & & \\%2
 & & & &0&9&8&8&8&8&8&8&8&9&0& & & & \\%3
 & & &0&9&8&8&7&7&7&7&7&8&8&9&0& & & \\%4
 & &0&9&8&8&7&6&6&6&6&6&7&8&8&9&0& & \\%5
 &0&9&8&8&7&6&5&5&5&5&5&6&7&8&8&9&0& \\%6
0&9&8&8&7&6&5&5&4&4&4&5&5&6&7&8&8&9&0\\%7
0&9&8&7&6&5&5&4&3&3&3&4&5&5&6&7&8&9&0\\%8
0&9&8&7&6&5&4&3&2&2&2&3&4&5&6&7&8&9&0\\%9
0&9&8&7&6&5&4&3&2&1&2&3&4&5&6&7&8&9&0\\%0
0&9&8&7&6&5&4&3&2&2&2&3&4&5&6&7&8&9&0\\%1
0&9&8&7&6&5&5&4&3&3&3&4&5&5&6&7&8&9&0\\%2
0&9&8&8&7&6&5&5&4&4&4&5&5&6&7&8&8&9&0\\%3
 &0&9&8&8&7&6&5&5&5&5&5&6&7&8&8&9&0& \\%4
 & &0&9&8&8&7&6&6&6&6&6&7&8&8&9&0& & \\%5
 & & &0&9&8&8&7&7&7&7&7&8&8&9&0& & & \\%6
 & & & &0&9&8&8&8&8&8&8&8&9&0& & & & \\%7
 & & & & &0&9&9&9&9&9&9&9&0& & & & & \\%7
 & & & & & &0&0&0&0&0&0&0& & & & & & \\%7
\end{array},
\end{gather*}
}}

\begin{figure}[htbp]
 \begin{tabular}{cc}
  \begin{minipage}[t]{0.45\hsize}
   \begin{center}
    \includegraphics[height=\hsize, clip, angle=270]{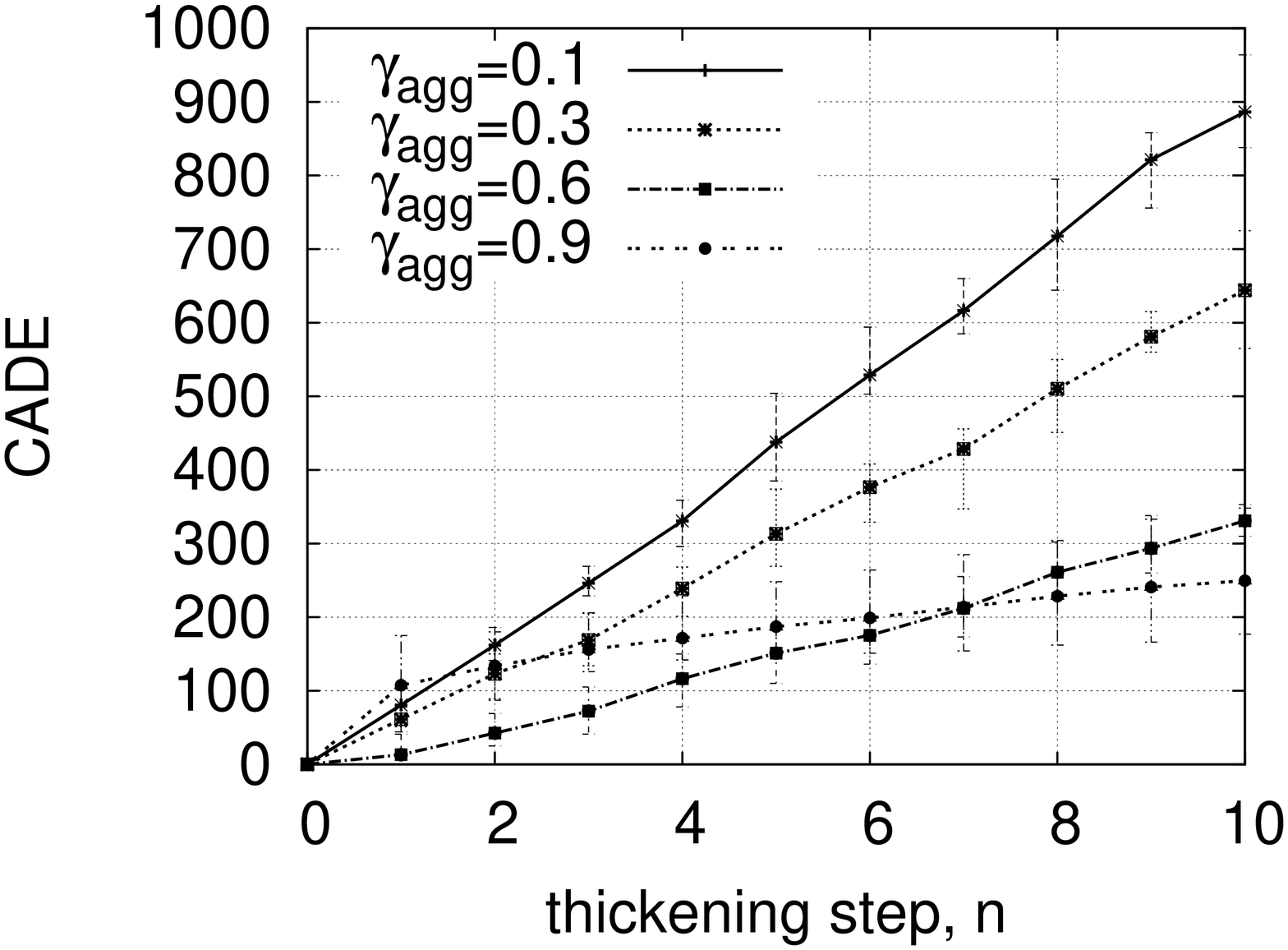}
    (a)
   \end{center}
   \begin{center}
    \includegraphics[height=\hsize, clip, angle=270]{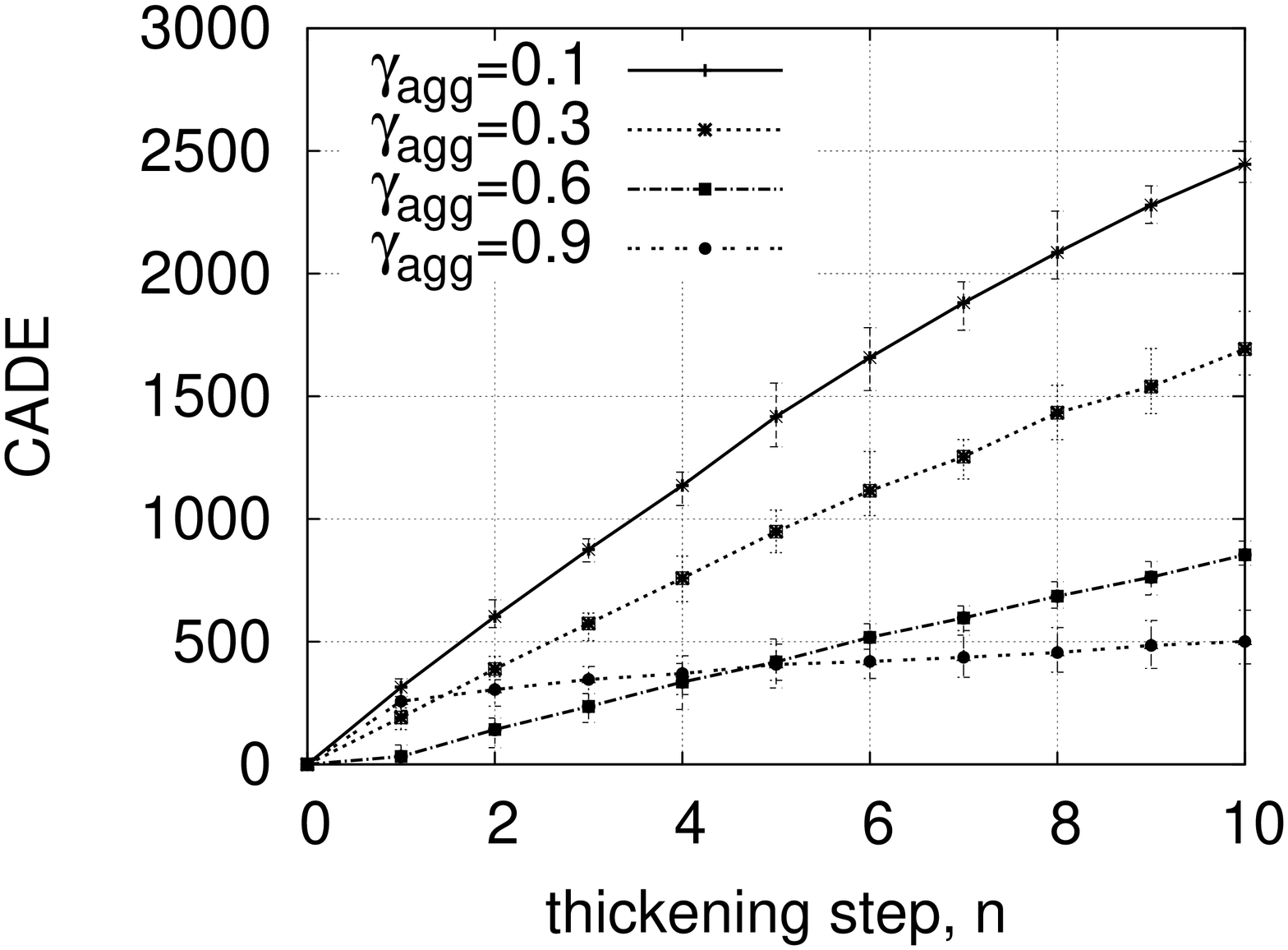}
    (c)
   \end{center}
  \end{minipage} 
  \begin{minipage}[t]{0.45\hsize}
   \begin{center}
    \includegraphics[height=\hsize, clip, angle=270]{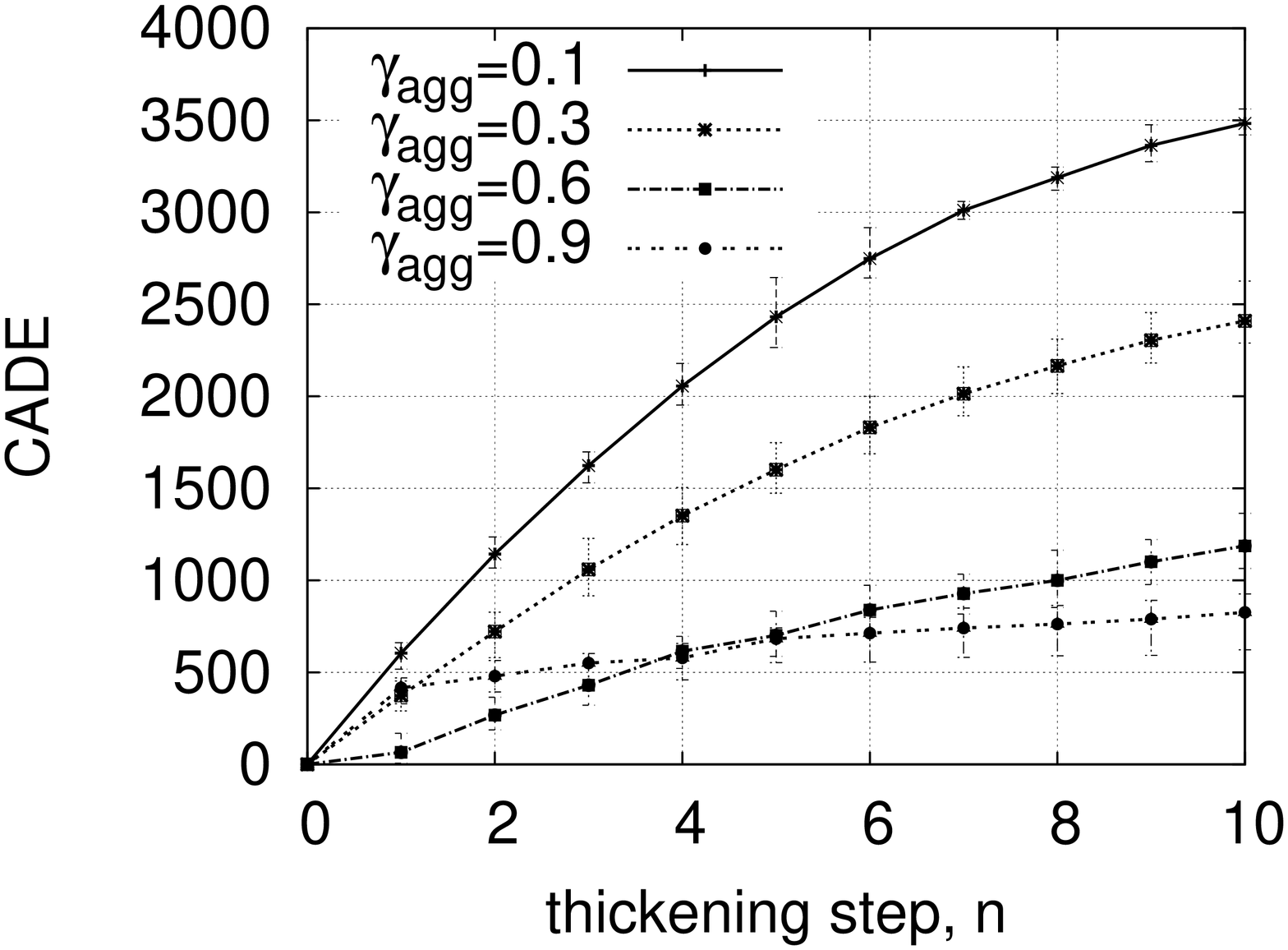}
    (b)
   \end{center}
   \begin{center}
    \includegraphics[height=\hsize, clip, angle=270]{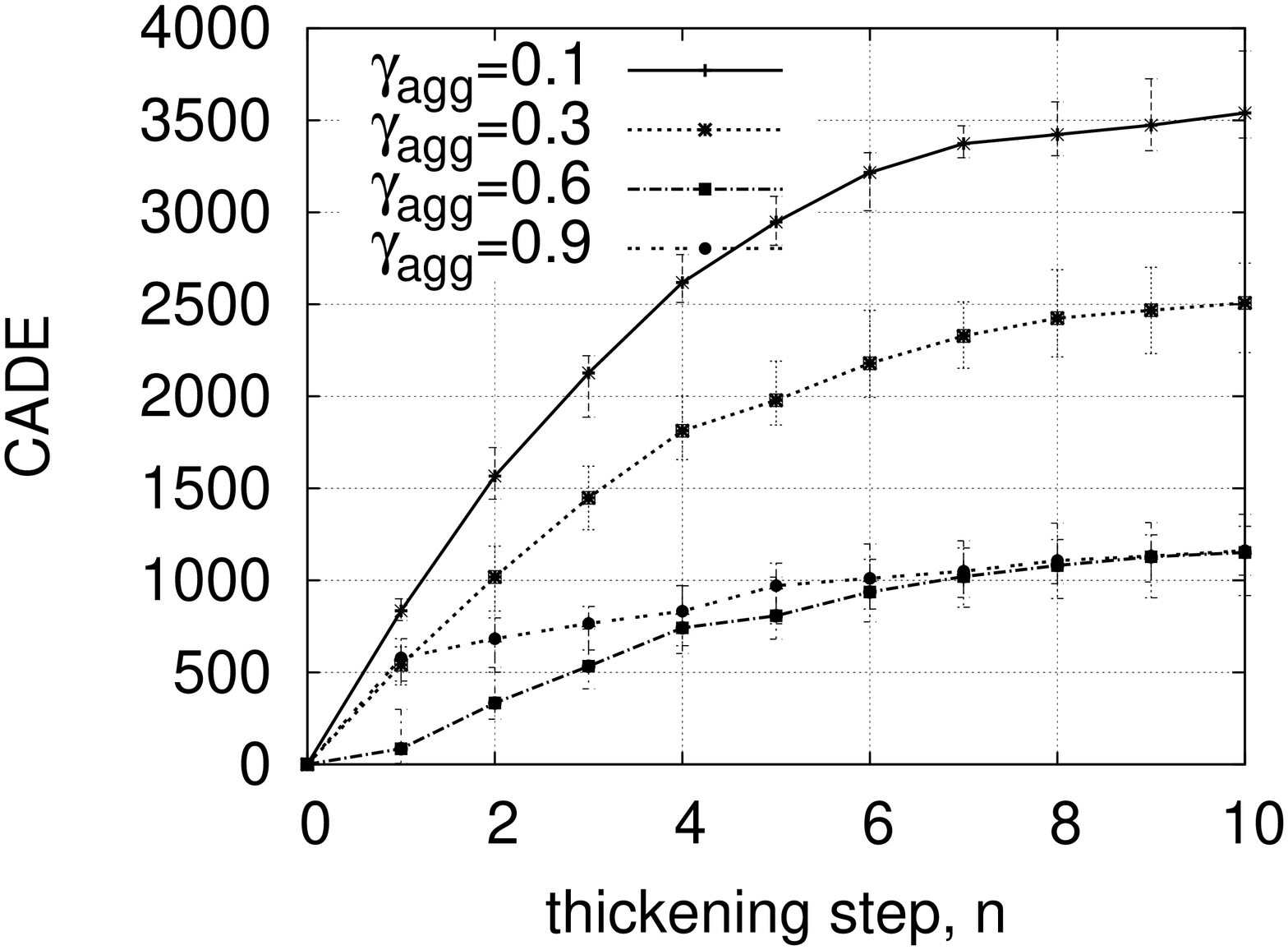}
    (d)
   \end{center}
  \end{minipage} 
 \end{tabular}
\caption{\label{fig:CADEvsTh}
The CADE   
$\hcE(\cM_{p,\gamma_\agg}; n a, 0)$ of the
$n$-th thickening step for
each $\gamma_\agg$;
Those of  the volume fraction $p$
$=0.1, 0.2, 0.3$ and $0.4$ are
illustrated in (a), (b), (c), and (d) respectively.
}
\end{figure}

We computed the ten pictures for each $p=0.1,0.2,0.3,0.4$
and $\gamma_\agg=0.0, 0.3, 0.6, 0.9$ with
ten random seeds. 
Figure \ref{fig:CADEvsTh} shows the CADE, 
$\hcE(\cM_{p,\gamma_\agg}; n a, 0)$, for the
$n$-th thickening step for each $p$.

\begin{figure}[htbp]
 \begin{tabular}{cc}
  \begin{minipage}[t]{0.45\hsize}
   \begin{center}
    \includegraphics[height=\hsize, clip, angle=270]{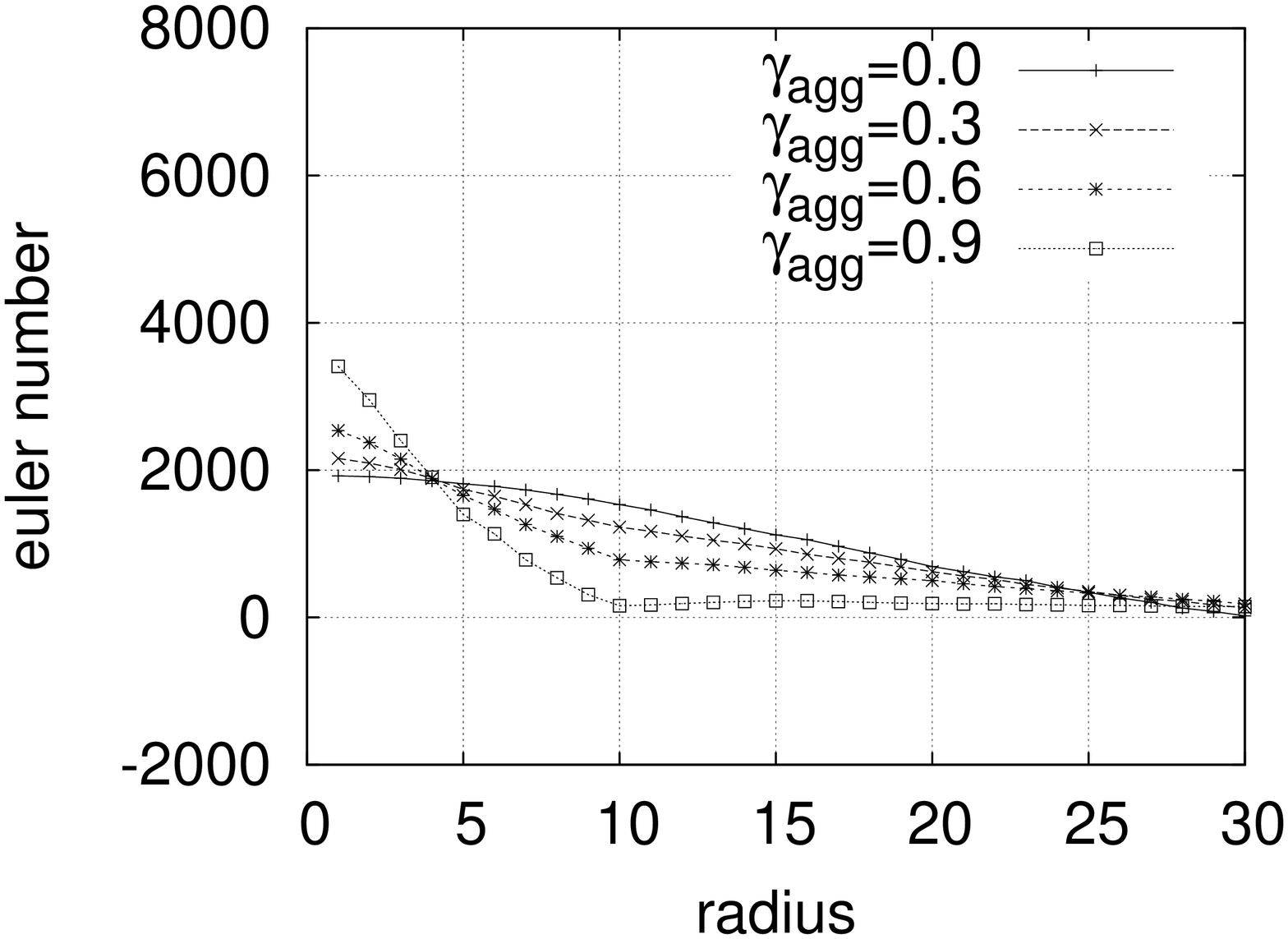}
    (a)
   \end{center}
   \begin{center}
    \includegraphics[height=\hsize, clip, angle=270]{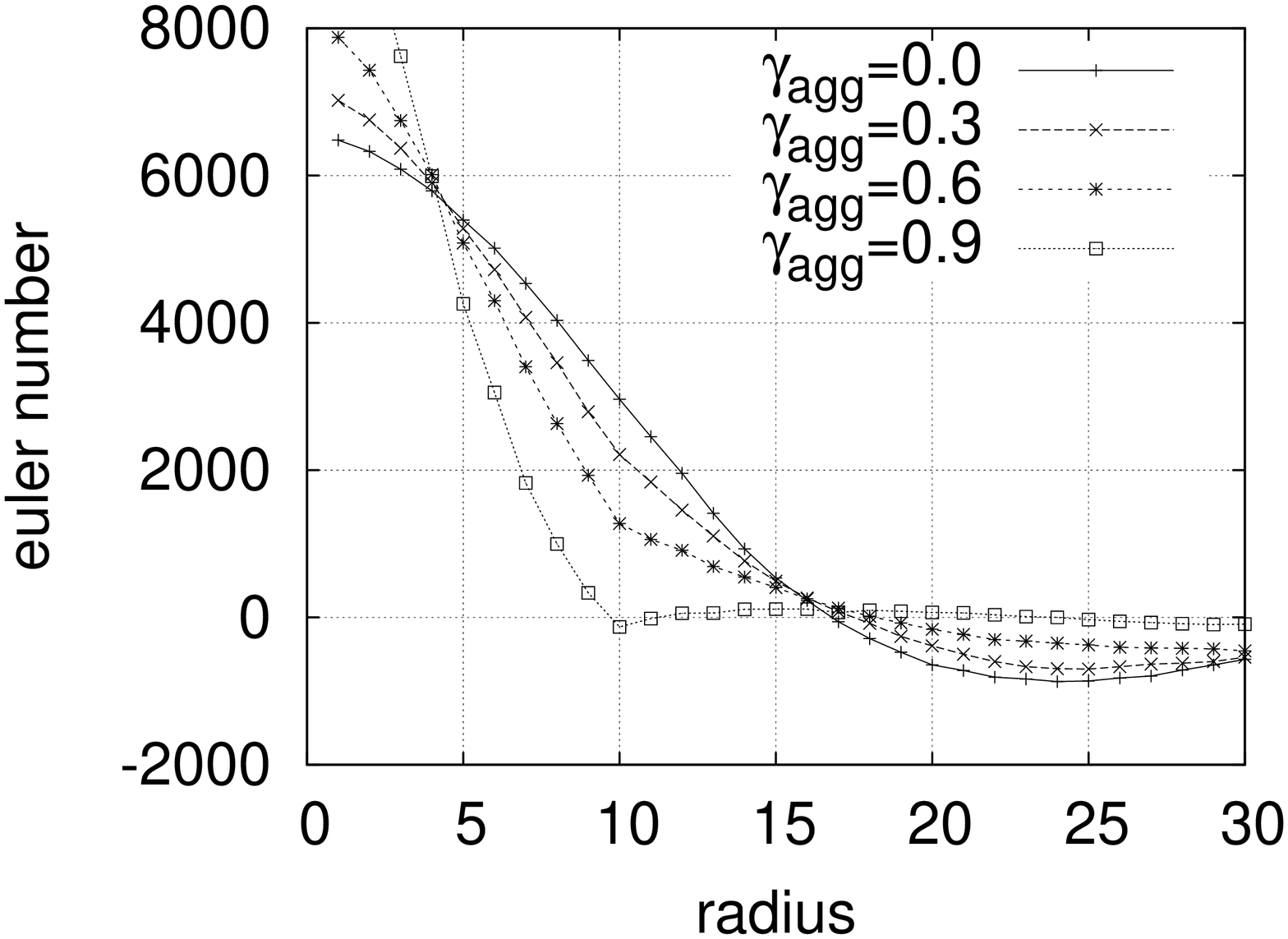}
    (c)
   \end{center}
  \end{minipage} 
  \begin{minipage}[t]{0.45\hsize}
   \begin{center}
    \includegraphics[height=\hsize, clip, angle=270]{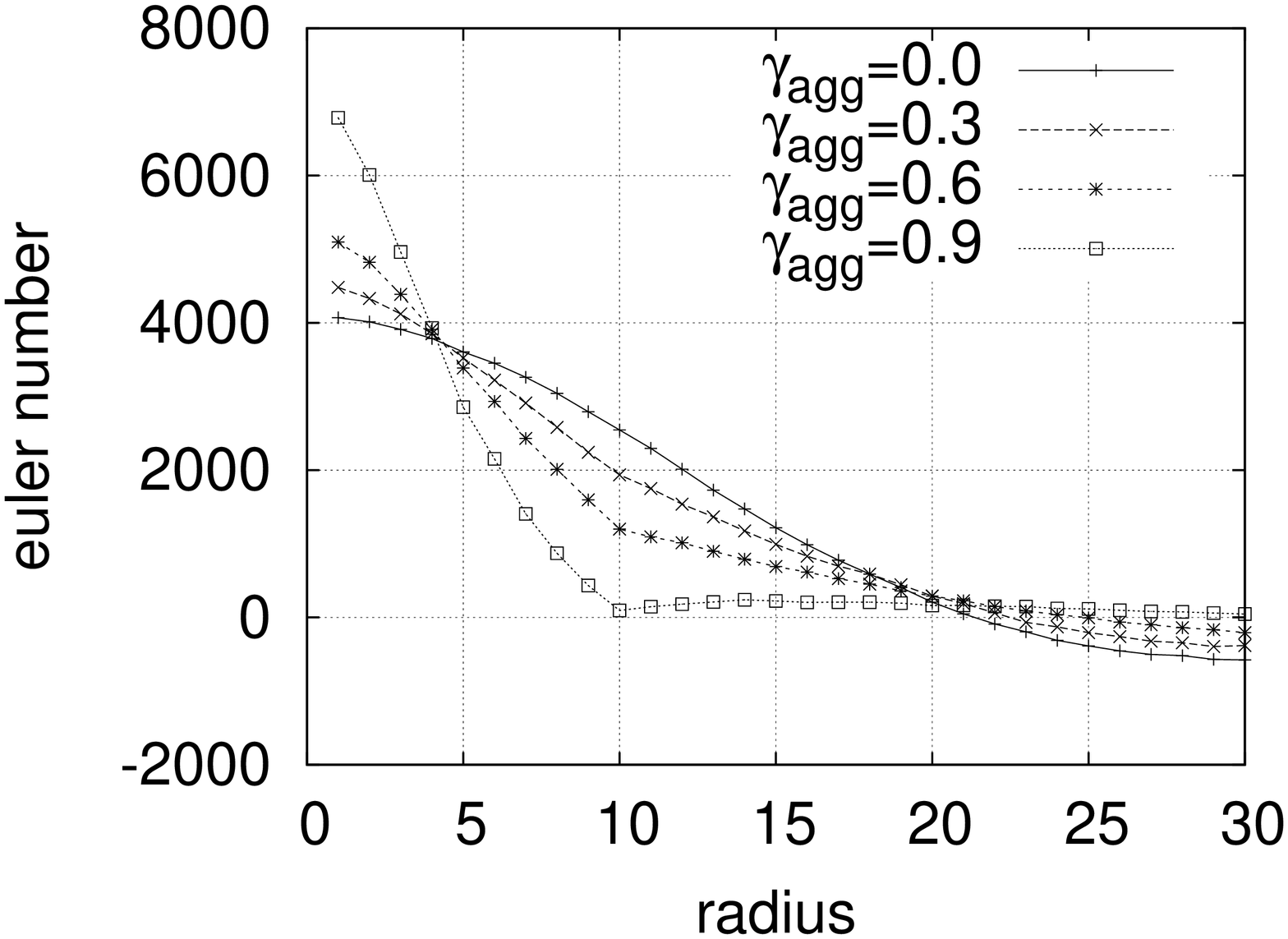}
    (b)
   \end{center}
   \begin{center}
    \includegraphics[height=\hsize, clip, angle=270]{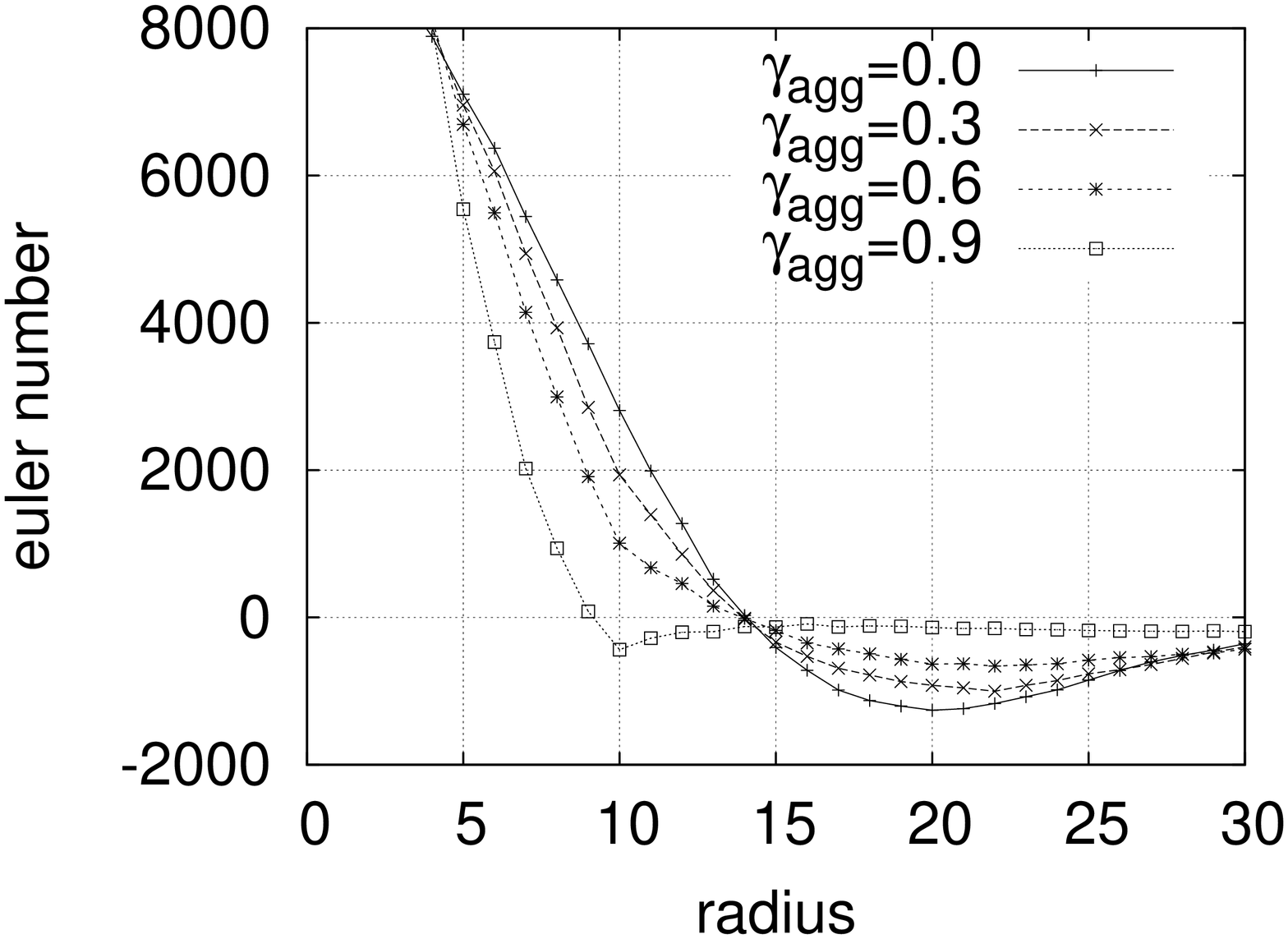}
    (d)
   \end{center}
  \end{minipage} 
 \end{tabular}
\caption{\label{fig:EulerPP}
The Euler number vs the radius $\rho$ as the point
pattern of $\cM_{\gamma_\agg,p,i_S}$:
 (a), (b), (c), and (d) 
illustrate the Euler numbers of the volume fraction $p$
$=0.1, 0.2, 0.3$ and $0.4$ respectively.
}
\end{figure}

On the other hand, 
though it is 
difficult to identify the center points of the particles
for given pictures,
especially of the agglomerated case
as shown in 
images (b) and (c) of
Figures \ref{fig:2Dview0.2} and \ref{fig:2Dview0.4},
we know the data of the center points of the particles.
Thus
we can use the techniques of the statistical analysis
for the spatial point patterns.
Figure \ref{fig:EulerPP} %\ref{fig:EulerPP}
displays the global distribution of the
Euler numbers of different radius of a seed
by using the software provided in \cite[p.204]{IPSS}\footnote{
\texttt{http://www.maths.jyu.fi/~penttine/ppstatistics.}}.

Since our radius is 10 and our agglomeration algorithm
is characterized by the radius, the behavior
of the distribution 
in Figure \ref{fig:EulerPP} 
strongly depends on the regions
$\rho > 10$ and $\rho \le 10$.
Figure \ref{fig:CADEvsTh} correspond to $(10,20]$ region and thus
it implies that our improved thickening algorithm works well
except the first thickening step of the $\gamma_\agg = 0.9$ case.

Since the agglomeration in our algorithm means that the number of
agglomerated particles is larger than the uniform
randomness $\gamma_\agg = 0$. 
The variation of the Euler number is related to 
the deformation in which disjoint clusters
connect due to the thickening. 
Agglomeration means that the number of the disjoint clusters
is less than that of uniform randomness.
The variation of
the Euler number for the increasing of the 
radius $\rho>10$ in Figure \ref{fig:EulerPP} 
is suppressed for large $\gamma_\agg$.
Hence the dependence in Figure \ref{fig:CADEvsTh} is very natural
except the first thickening step of the $\gamma_\agg = 0.9$ case.

Further in the image processing procedure, we must pay attention
to the digitalized errors. We should recognize that
the first step contains some digitalized errors because
the behavior in Figure \ref{fig:CADEvsTh}
is contradict with that in 
Figure \ref{fig:EulerPP}; the behavior of the 
curves of $\gamma_\agg = 0.9$ in Figure \ref{fig:EulerPP}
are very mild over $(10,20)$ whereas the first steps
of $\gamma_\agg = 0.9$ in Figure \ref{fig:CADEvsTh} 
rapidly increase.

Hence we are concerned with 
$\hcE(\cM_{p,\gamma_\agg}; \rho, a)$.

%In order to find these behaviors more precisely,
%we consider the dependence of the CADE 
%$\hcE(\cM_{p,\gamma_\agg}; \rho, a)$
%on the agglomeration parameter $\gamma_\agg$.

\begin{figure}[htbp]
 \begin{tabular}{cc}
  \begin{minipage}[t]{0.45\hsize}
   \begin{center}
    \includegraphics[height=\hsize, clip, angle=270]{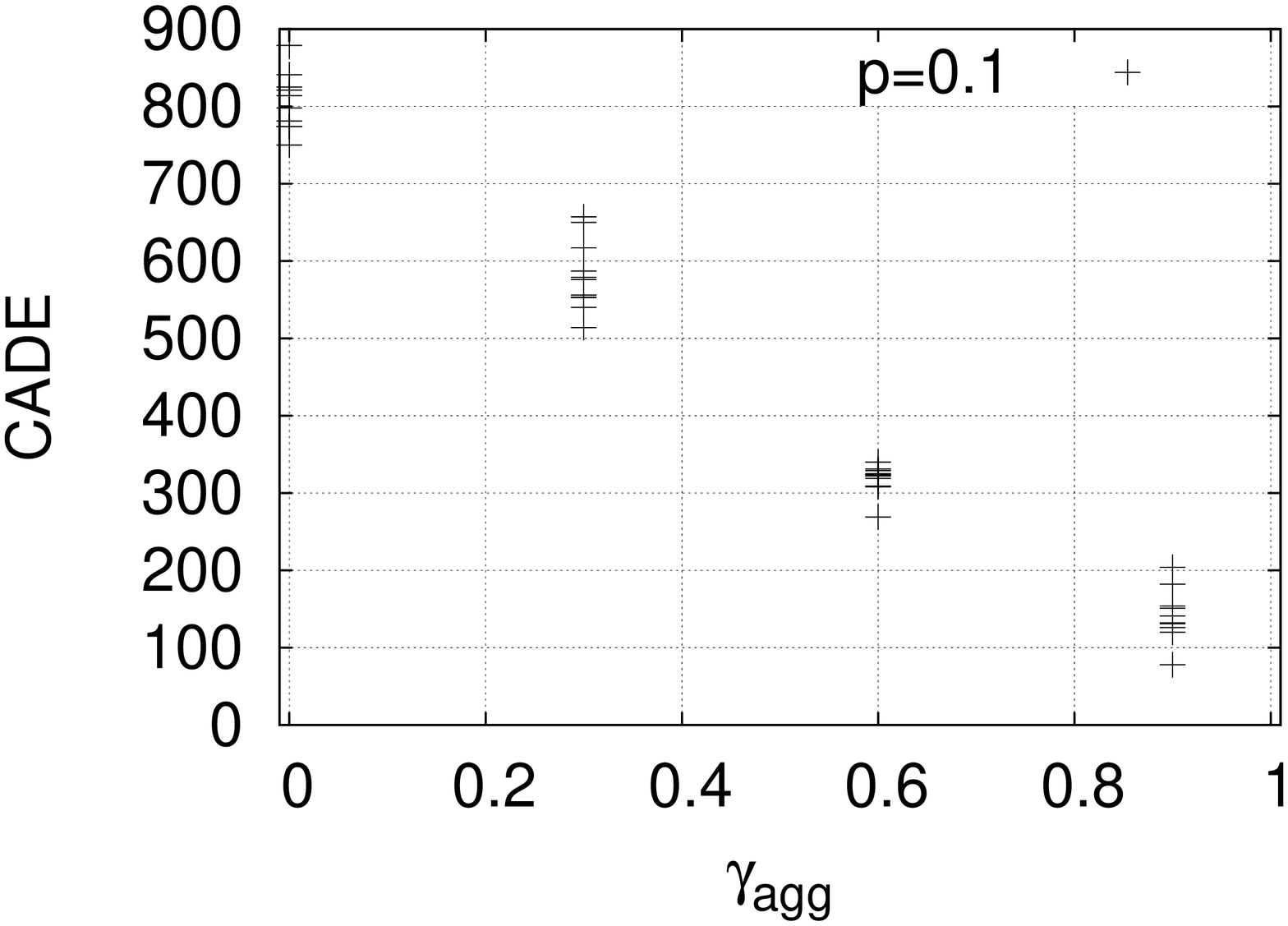}
    (a)
   \end{center}
   \begin{center}
    \includegraphics[height=\hsize, clip, angle=270]{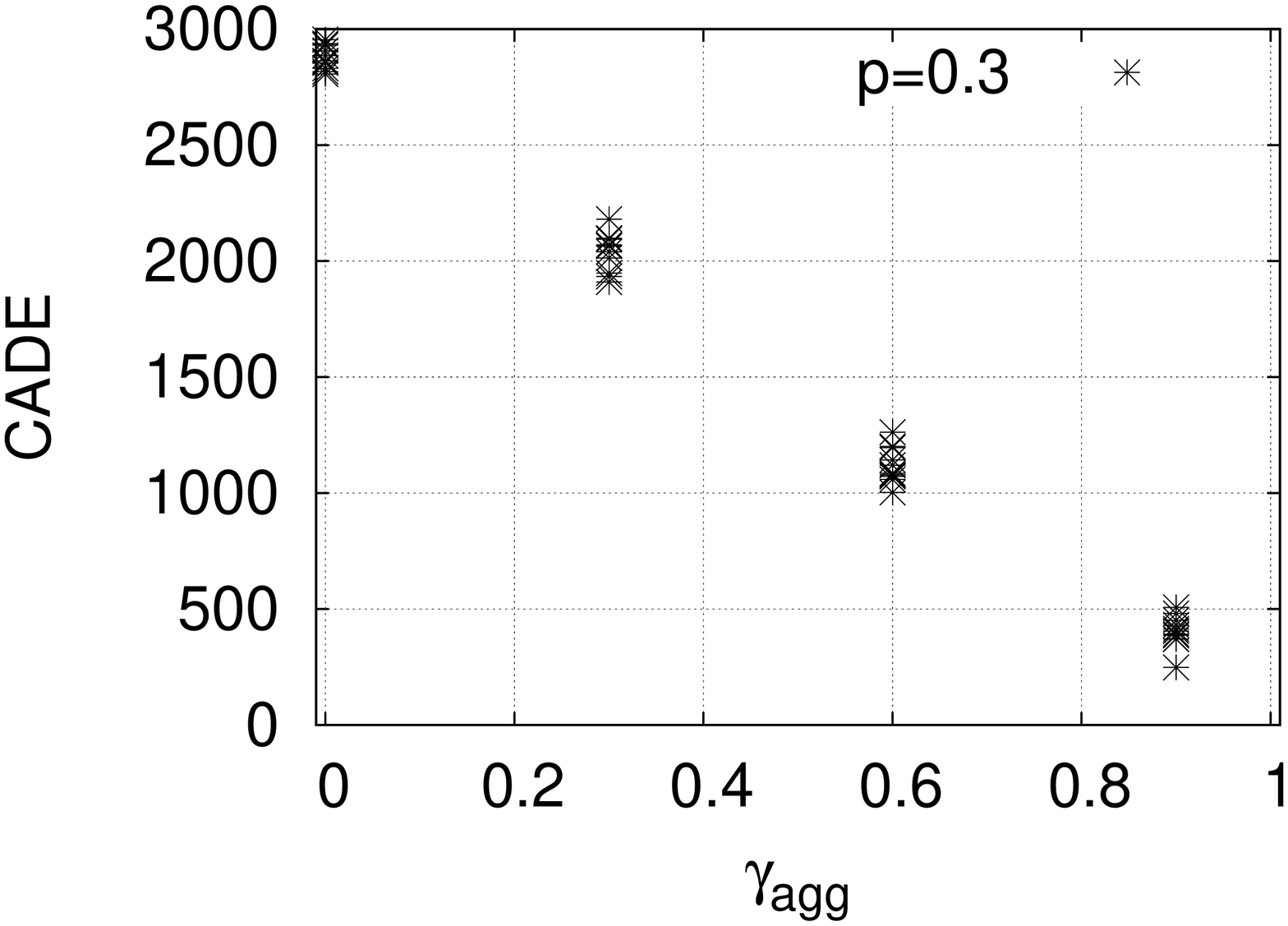}
    (c)
   \end{center}
  \end{minipage} 
  \begin{minipage}[t]{0.45\hsize}
   \begin{center}
    \includegraphics[height=\hsize, clip, angle=270]{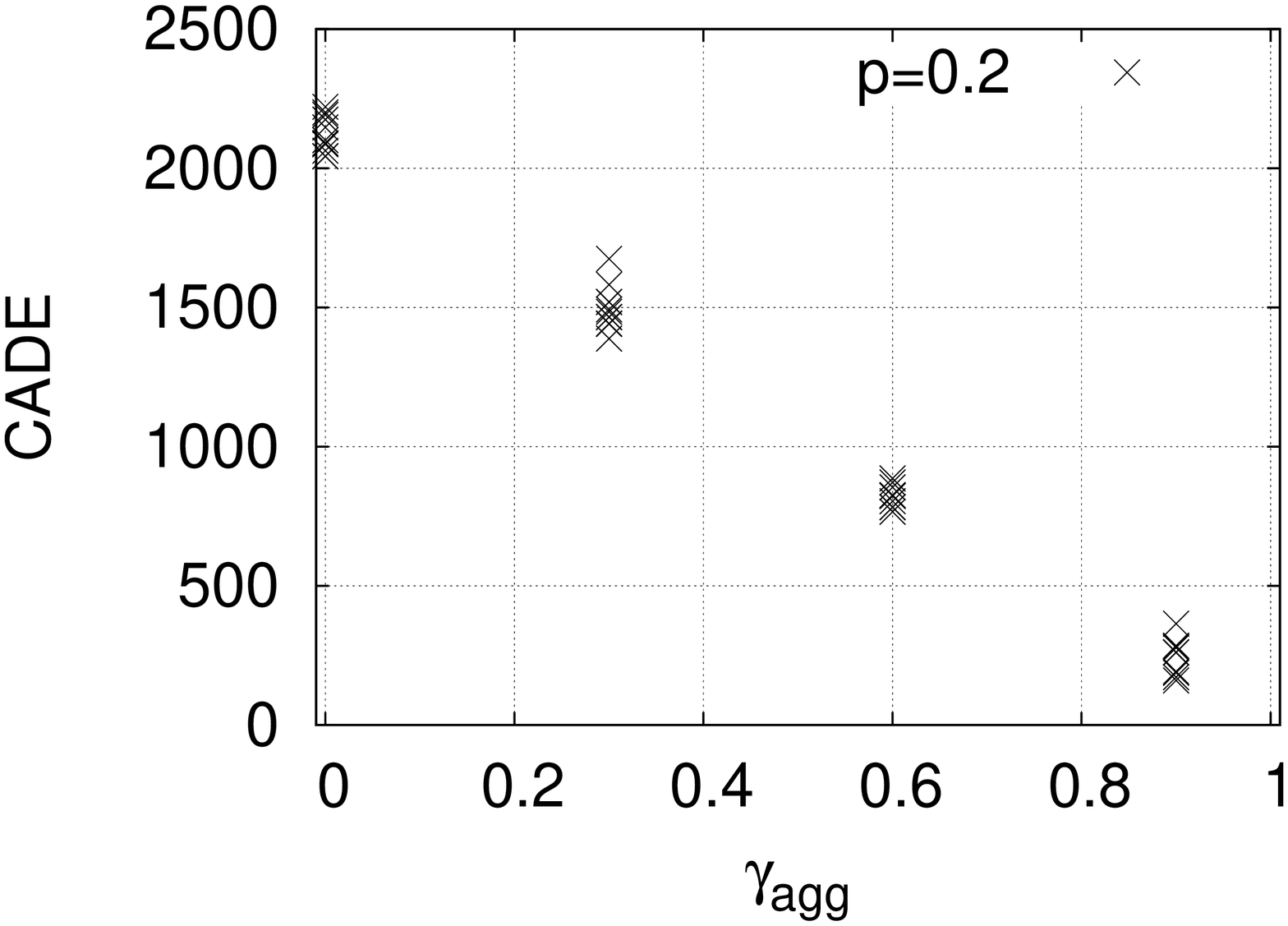}
    (b)
   \end{center}
   \begin{center}
    \includegraphics[height=\hsize, clip, angle=270]{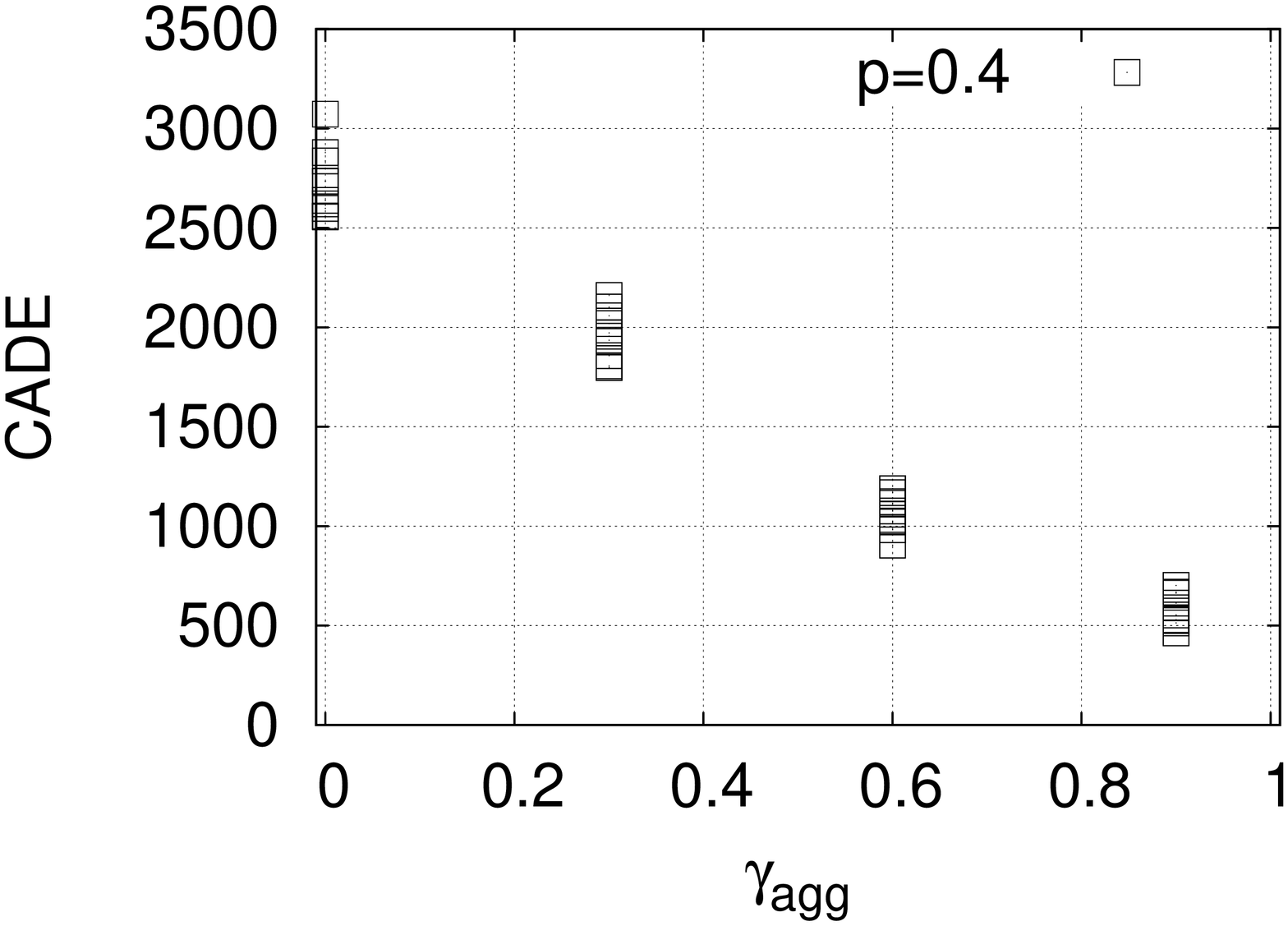}
    (d)
   \end{center}
  \end{minipage} 
 \end{tabular}
\caption{\label{fig:CADEgamma}
The CADE, $\hcE(\cM_{p,\gamma_\agg}; \rho, a)$, vs $\gamma_\agg$:
 (a), (b), (c), and (d) 
display the states of the volume fraction $p$
$=0.1, 0.2, 0.3$ and $0.4$ respectively.
}
\end{figure}

\begin{figure}[htbp]
 \begin{tabular}{cc}
  \begin{minipage}[t]{0.45\hsize}
   \begin{center}
    \includegraphics[height=\hsize, clip, angle=270]{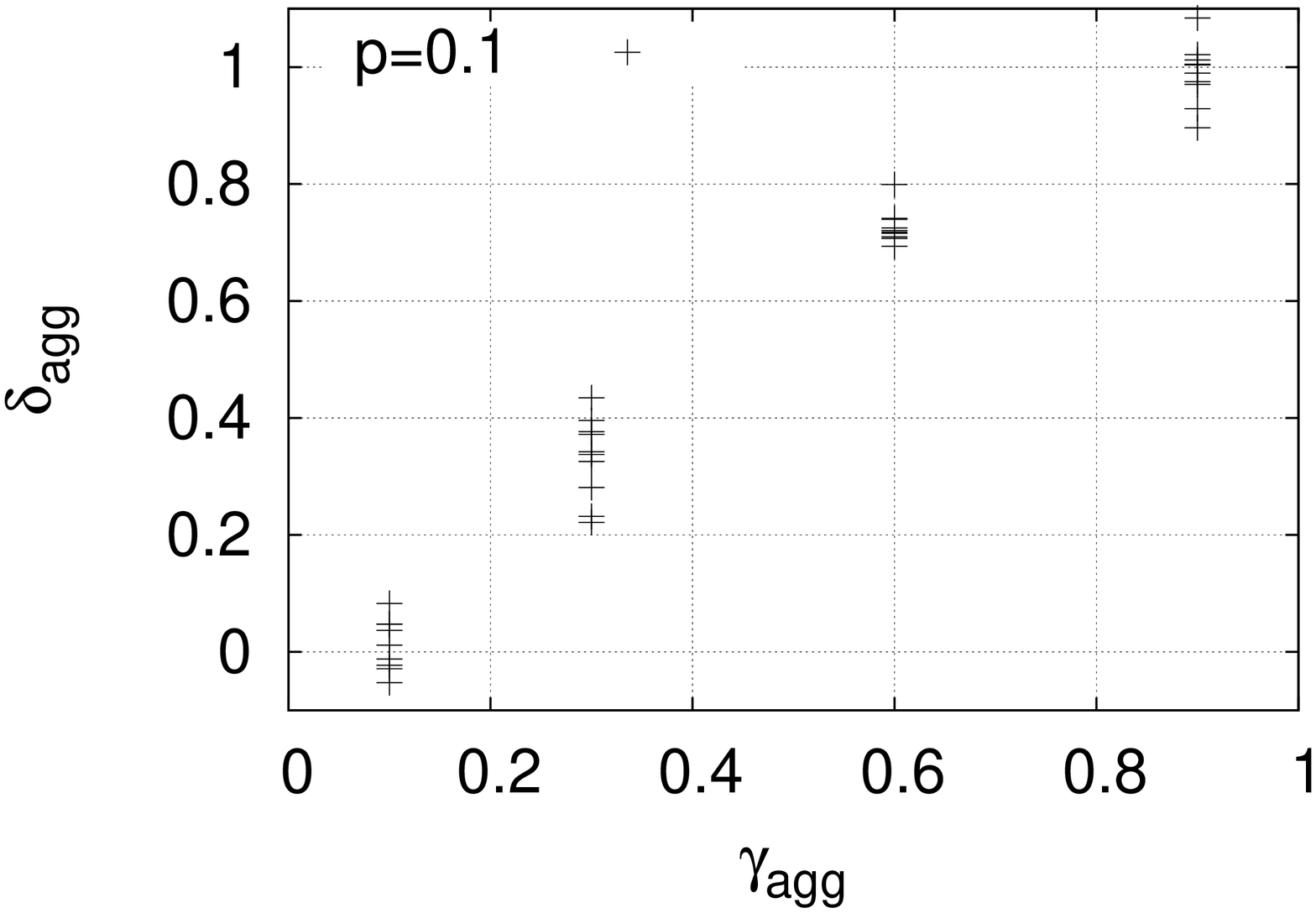}
    (a)
   \end{center}
   \begin{center}
    \includegraphics[height=\hsize, clip, angle=270]{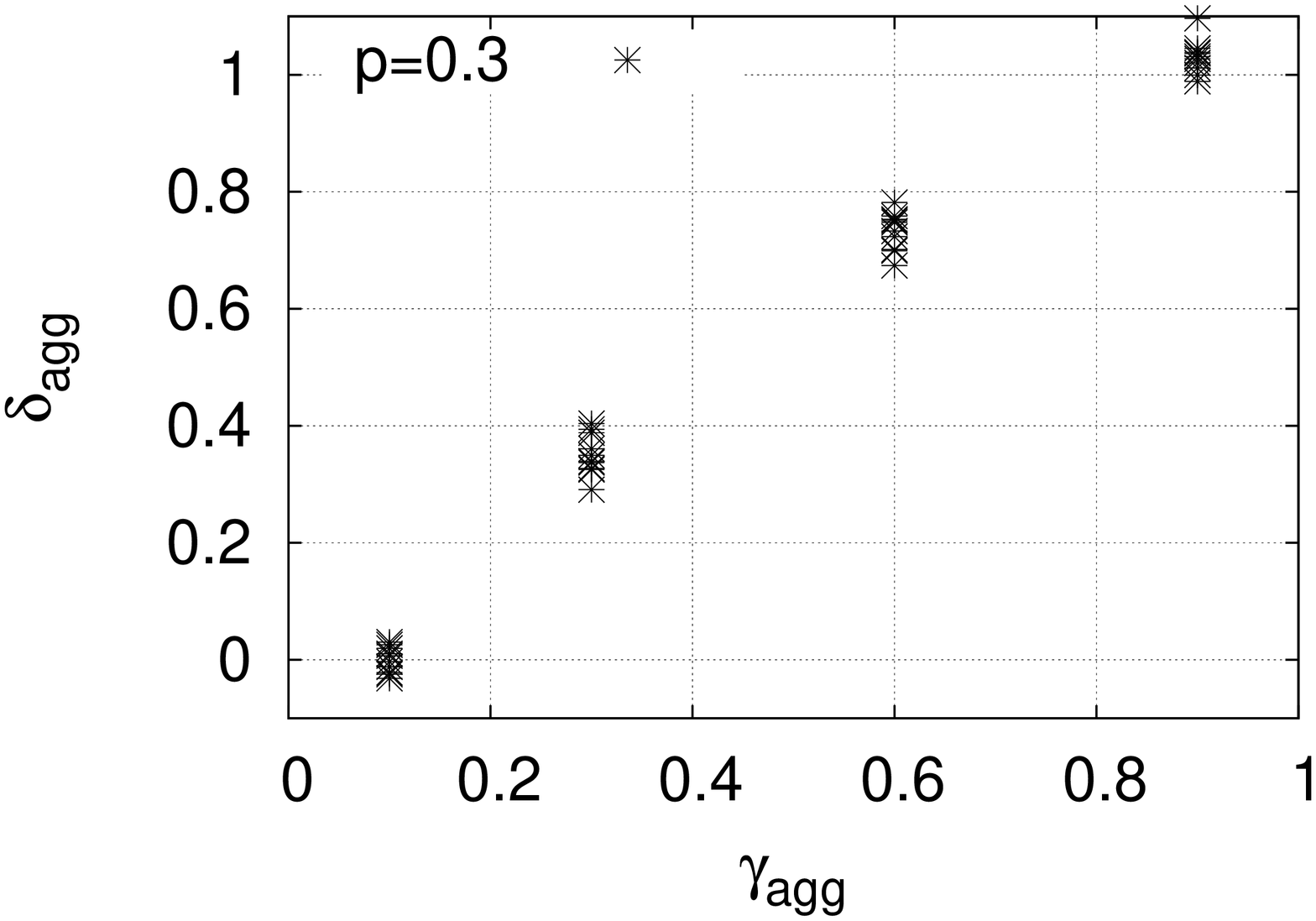}
    (c)
   \end{center}
  \end{minipage} 
  \begin{minipage}[t]{0.45\hsize}
   \begin{center}
    \includegraphics[height=\hsize, clip, angle=270]{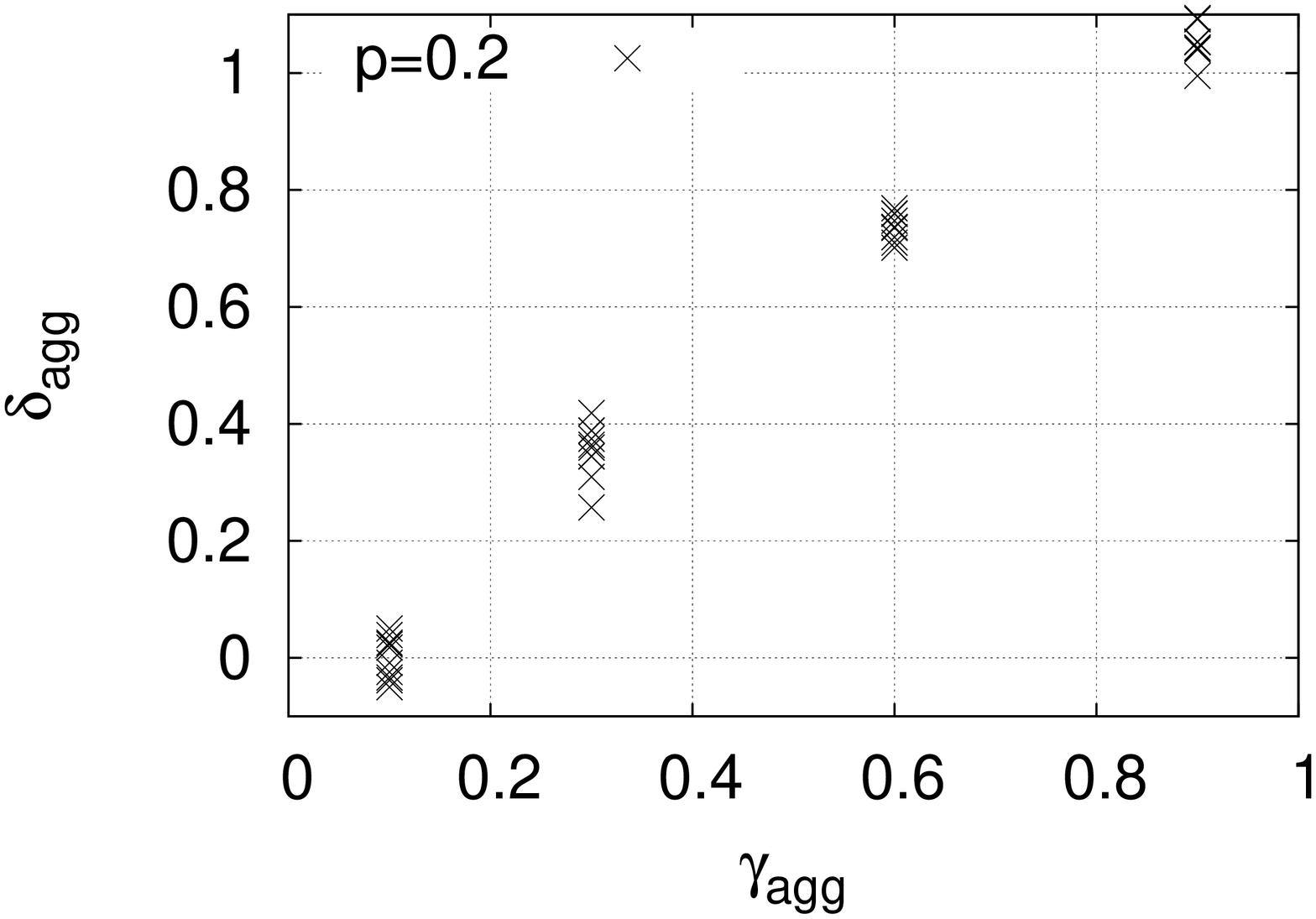}
    (b)
   \end{center}
   \begin{center}
    \includegraphics[height=\hsize, clip, angle=270]{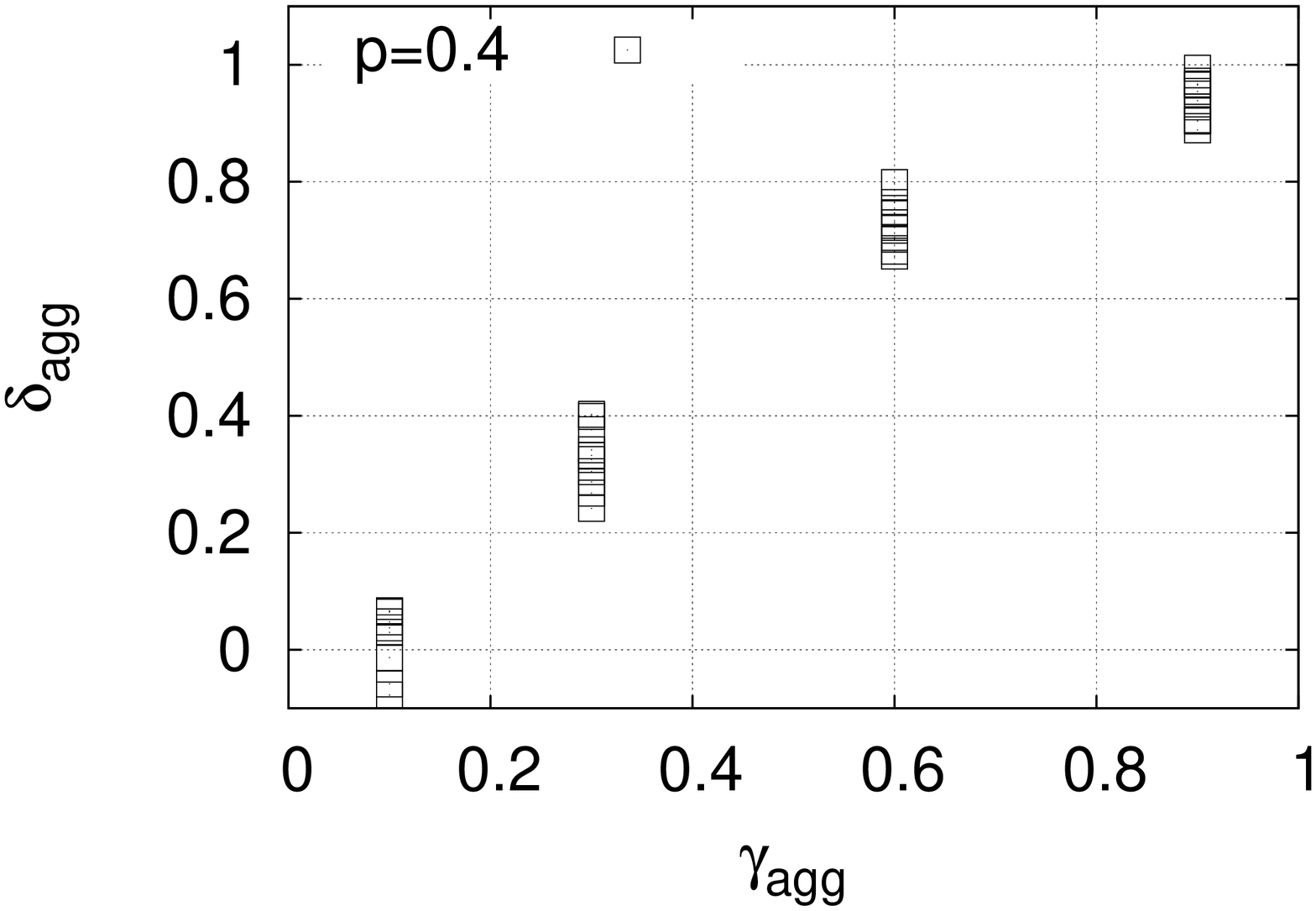}
    (d)
   \end{center}
  \end{minipage} 
 \end{tabular}
\caption{\label{fig:deltagamma}
The agglomeration index $\delta_\agg$ 
vs $\gamma_\agg$:
 (a), (b), (c), and (d) 
display the states of the volume fraction $p$
$=0.1, 0.2, 0.3$ and $0.4$ respectively.
}
\end{figure}

Table \ref{tbl:CADEvsgamma} and Figure \ref{fig:CADEgamma} 
 show the dependence of the CADE,
$\hcE(\cM_{p,\gamma_\agg}; \rho, a)$,
on the agglomeration parameter $\gamma_\agg$ for 
each $p$.
They exhibit the negative correlations.
The agglomeration means that the number of
agglomerated particles is larger than the uniform
randomness $\gamma_\agg = 0$ as mentioned above.
Since the agglomeration prevents the topological
changes on the thickening, we have
the negative correlation in Figure \ref{fig:CADEgamma}.

\begin{table}[htbp]
{\small{
\caption{CADE vs $\gamma_\agg$:}
\label{tbl:CADEvsgamma}
\begin{center}
\begin{tabular}{|c|c|c|c|c|c|c|}
\hline
\multicolumn{1}{|c|}{$p$} &
\multicolumn{3}{|c|}{0.1}&
\multicolumn{3}{|c|}{0.2}\\
\hline
$\gamma_\agg$ &Ave& Max&Min & Ave&Max&Min \\
\hline
0  &805.7 &879 &750 &2131.9 &2221 &2043 \\
0.3&582.9 &657 &514 &1501.2 &1675 &1388 \\
0.6&317.6 &340 &269 &821.9  &886  &766  \\
0.9&141.9 &204 &78  &244.0  &364  &160  \\

\hline
\hline
\multicolumn{1}{|c|}{$p$} &
\multicolumn{3}{|c|}{0.3}&
\multicolumn{3}{|c|}{0.4}\\
\hline
$\gamma_\agg$ &Ave& Max&Min & Ave&Max&Min \\
\hline
0   &2878.5 &2954 &2806 &2705.3 &3072 &2555 \\
0.3 &2035.2 &2181 &1910 &1966.3 &2160 &1798 \\
0.6 &1121.9 &1262 &1003 &1066.6 &1188 &905  \\
0.9 &408.8  &507  &248  &582.1  &702  &464  \\
\hline
\end{tabular}
\end{center}
}}
\end{table}

We, now, define  the 
 the average $\hcE_p$ 
of a ``standard pattern of volume fraction $p$" 
by the average of the CADE, $\hcE(\cM_{p,0,i_S};$ $ \rho, a)$, 
of the uniform random configuration,
 i.e., $\gamma_\agg = 0$ case,
$$
\hcE_p\equiv \hcE^{(\rho/a, 1)}_p:=
 \frac{1}{10}\sum_{i_S=1}^{10} \hcE(\cM_{0,p,i_S}; \rho,a).
$$
Thus we denote the agglomeration index
$\delta_\agg^{(\rho/a,1)}$ by $\delta_\agg$
as in (\ref{eq:delta}).
Further in order that $\delta_\agg$ corresponds to
$\gamma_\agg$, we chose $\alpha = 1.2$.

Figure \ref{fig:deltagamma} and 
Table \ref{tbl:deltavsgamma} show the relation between
$\delta_\agg$ and $\gamma_\agg$;
both show that $\gamma_\agg$ is correlated to $\delta_\agg$ and 
approximately recovers $\delta_\agg$
up to the statistical fluctuation.

\begin{table}[htbp]
{\small{
\caption{$\delta_\agg$ vs $\gamma_\agg$:}
\label{tbl:deltavsgamma}
\begin{center}
\begin{tabular}{|c|c|c|c|c|c|c|}
\hline
\multicolumn{1}{|c|}{$p$} &
\multicolumn{3}{|c|}{0.1}&
\multicolumn{3}{|c|}{0.2}\\
\hline
$\gamma_\agg$ &Ave& Max&Min & Ave&Max&Min \\
\hline
0  & 0.000 &0.083 &-0.109 &0.000 &0.050 &-0.050 \\
0.3& 0.332 &0.434 &0.221  &0.355 &0.419 &0.257  \\
0.6& 0.727 &0.799 &0.694  &0.737 &0.769 &0.701  \\
0.9& 0.989 &1.084 &0.896  &1.063 &1.110 &0.995  \\
\hline
\hline
\multicolumn{1}{|c|}{$p$} &
\multicolumn{3}{|c|}{0.3}&
\multicolumn{3}{|c|}{0.4}\\
\hline
$\gamma_\agg$ &Ave& Max&Min & Ave&Max&Min \\
\hline
0  & 0.000 &0.030 &-0.031 &0.000 &0.067 &-0.163 \\
0.3& 0.352 &0.404 &0.291  &0.328 &0.402 &0.242  \\
0.6& 0.732 &0.782 &0.674  &0.727 &0.799 &0.673  \\
0.9& 1.030 &1.097 &0.989  &0.942 &0.994 &0.889  \\
\hline
\end{tabular}
\end{center}
}}
\end{table}

\begin{figure}[htbp]
 \begin{tabular}{cc}
  \begin{minipage}[t]{0.5\hsize}
   \begin{center}
    \includegraphics[height=\hsize, clip, angle=270]{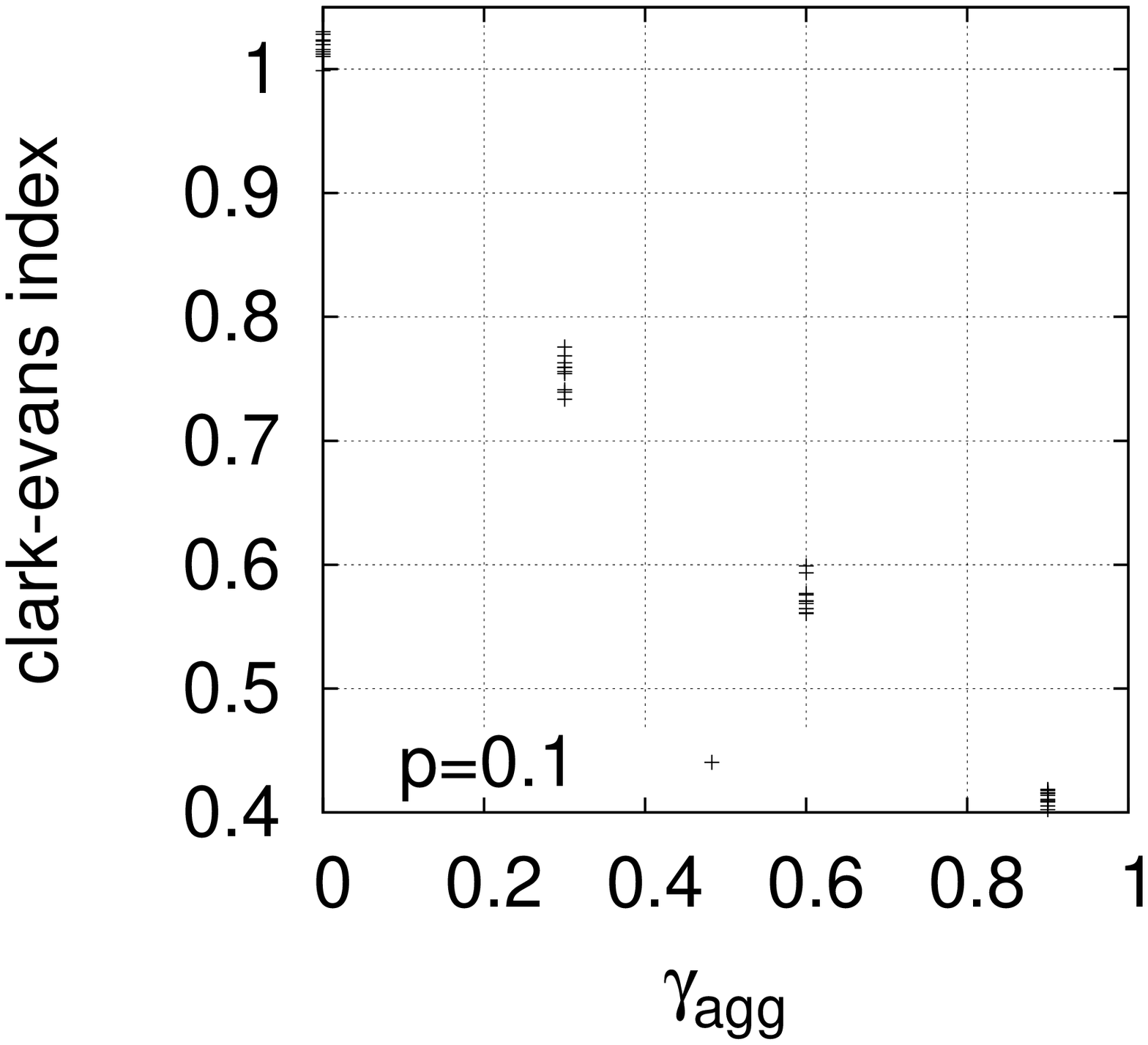}
    (a)
   \end{center}
   \begin{center}
    \includegraphics[height=\hsize, clip, angle=270]{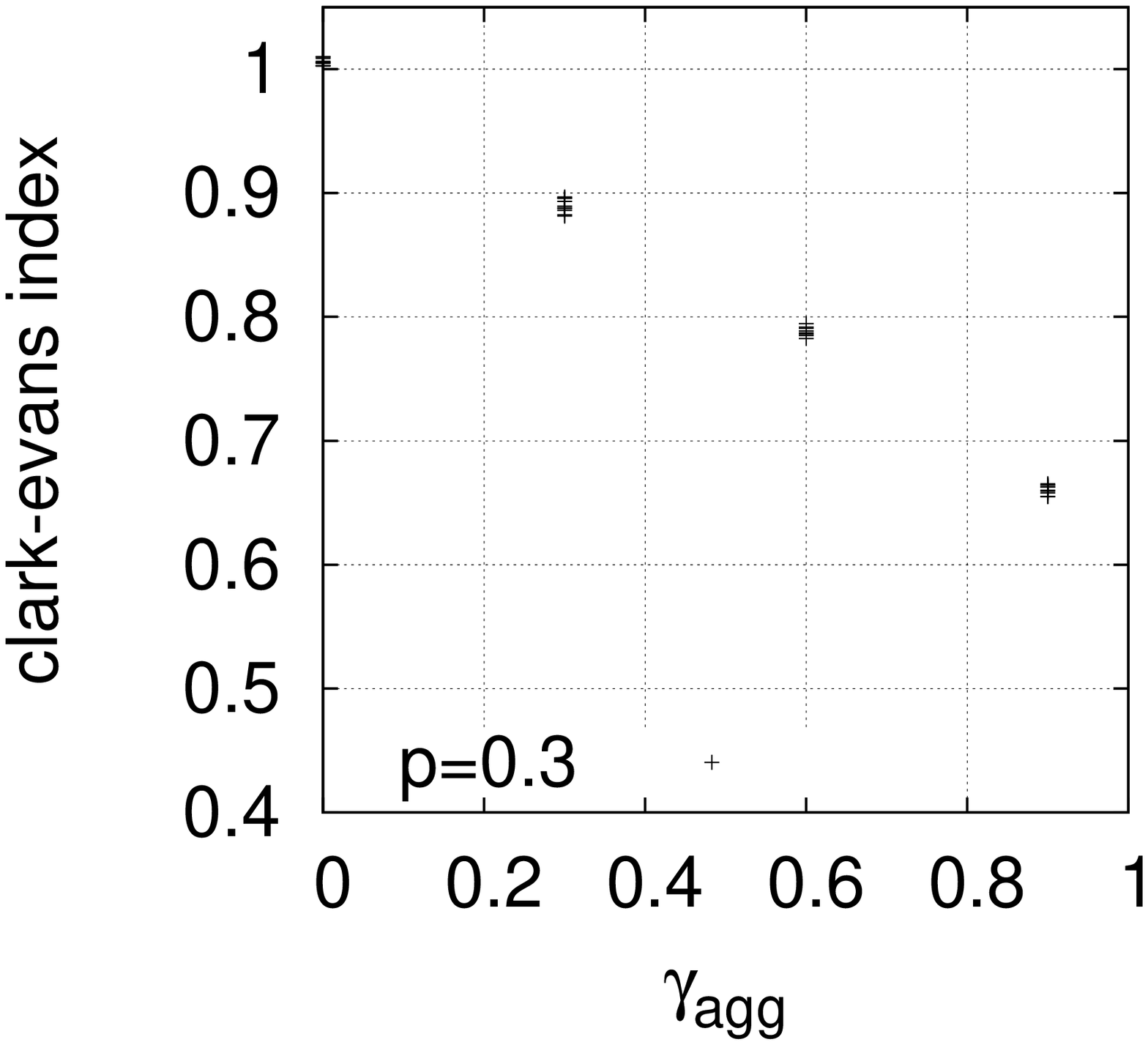}
    (c)
   \end{center}
  \end{minipage} 
  \begin{minipage}[t]{0.5\hsize}
   \begin{center}
    \includegraphics[height=\hsize, clip, angle=270]{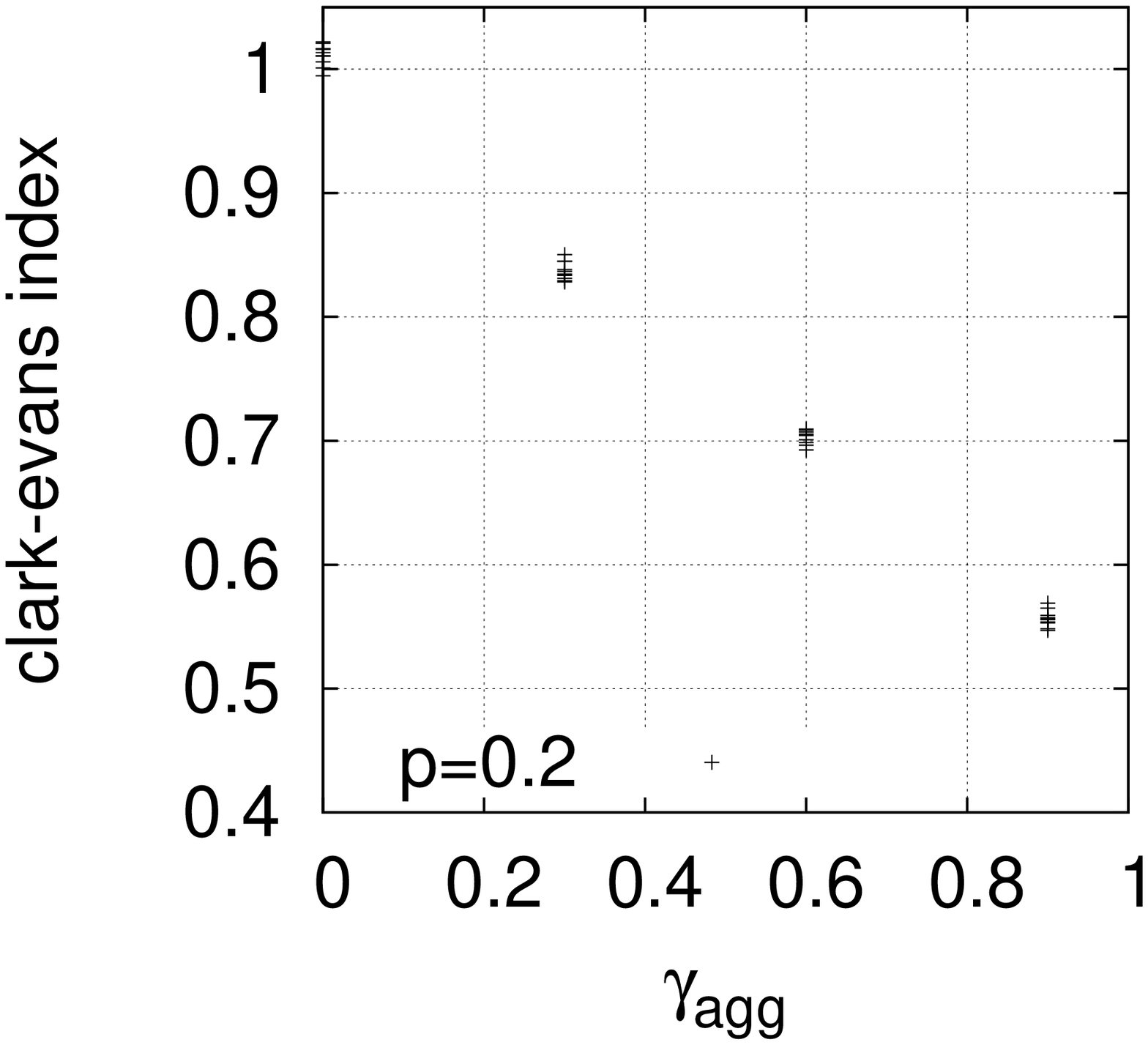}
    (b)
   \end{center}
   \begin{center}
    \includegraphics[height=\hsize, clip, angle=270]{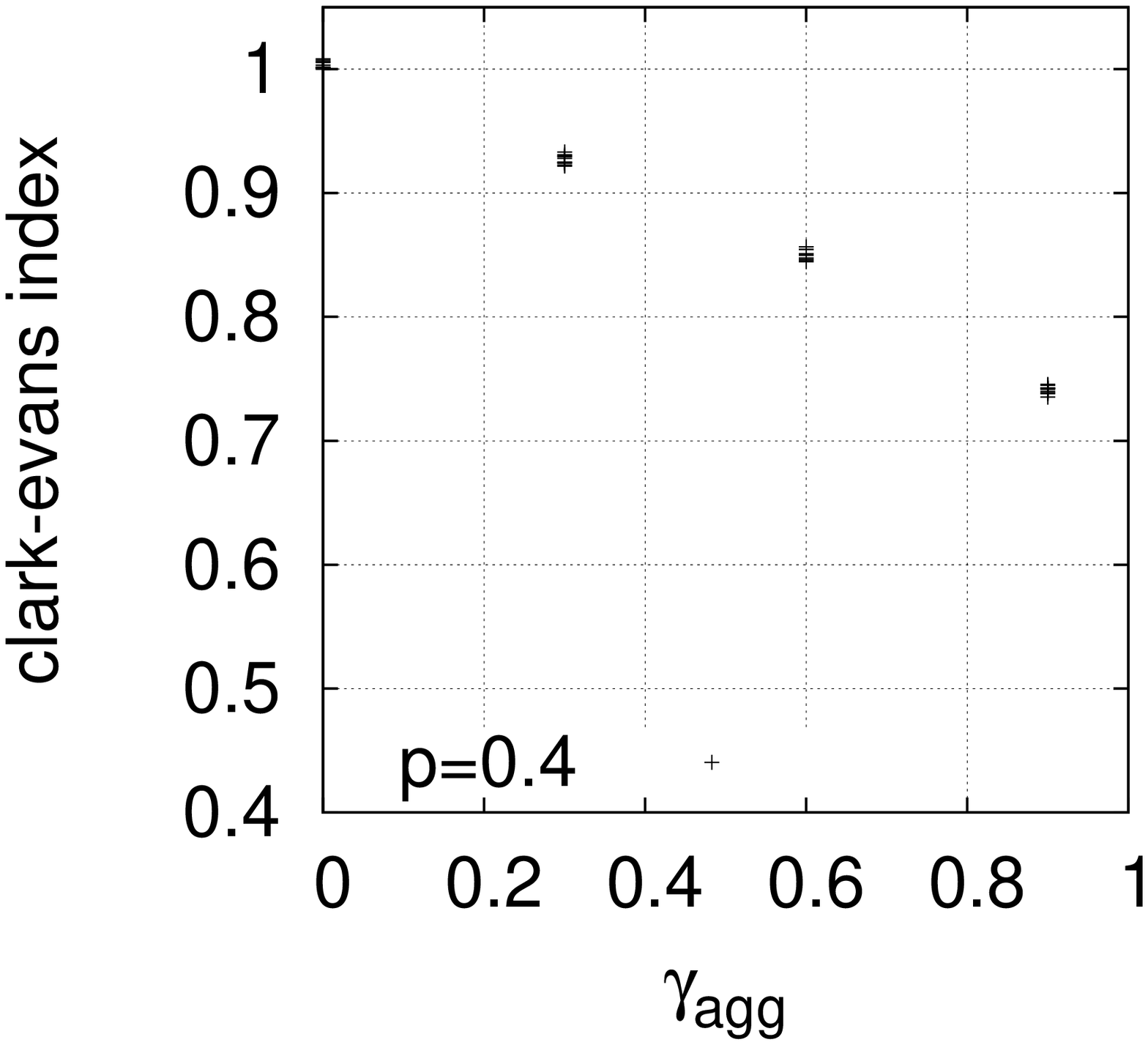}
    (d)
   \end{center}
  \end{minipage} 
 \end{tabular}
\caption{\label{fig:CEIvsgamma}
The Clark-Evans index vs $\gamma_\agg$:
 (a), (b), (c), and (d) 
display the states of the volume fraction $p$
$=0.1, 0.2, 0.3$ and $0.4$ respectively.
}
\end{figure}

\begin{figure}[htbp]
   \begin{center}
\includegraphics[height=\hsize, clip, angle=270]{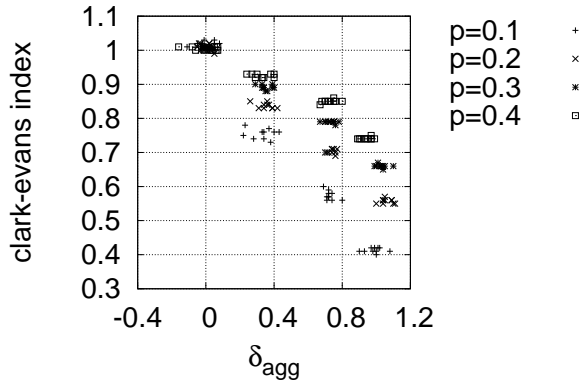}
   \end{center}
\caption{\label{fig:CEi}
The Clark-Evans index and $\delta_\agg$.
}
\end{figure}

In the  statistical analysis of the spatial
point patterns, the Clark-Evans index is a well-established
index which represents the agglomeration degree of
a given point pattern, though in general, it is very 
difficult to identify the center points of the 
particles for a given picture, such as 
images (b) and (c) of
Figures \ref{fig:2Dview0.2} and \ref{fig:2Dview0.4};
the problem is sometimes ill-posed for the cases.
Since we know the data of the 
center points of the particles of every $\cM_{\gamma_\agg, p, i_S}$, 
we illustrated the Clark-Evans index in  Table \ref{tbl:CEIvsgamma}
and Figure \ref{fig:CEIvsgamma},
which show that
the Clark-Evans index represents our
agglomeration parameter $\gamma_\agg$ well.
The correlation between the Clark-Evans index and
$\delta_\agg$ is displayed in Figure \ref{fig:CEi}.
It shows a good negative-correlation for each volume fraction
$p$.

\begin{table}[htbp]
{\small{
\caption{Clark Evans index vs $\gamma_\agg$:}
\label{tbl:CEIvsgamma}
\begin{center}
\begin{tabular}{|c|c|c|c|c|c|c|}
\hline
\multicolumn{1}{|c|}{$p$} &
\multicolumn{3}{|c|}{0.1}&
\multicolumn{3}{|c|}{0.2}\\
\hline
$\gamma_\agg$ &Ave& Max&Min & Ave&Max&Min \\
\hline
0&1.018 &1.030 &0.999 &1.011 &1.022 &0.995 \\
0.3&0.755 &0.776 &0.733 &0.837 &0.850 &0.828 \\
0.6&0.574 &0.599 &0.561 &0.703 &0.709 &0.693 \\
0.9&0.412 &0.419 &0.402 &0.556 &0.569 &0.547 \\
\hline
\hline
\multicolumn{1}{|c|}{$p$} &
\multicolumn{3}{|c|}{0.3}&
\multicolumn{3}{|c|}{0.4}\\
\hline
$\gamma_\agg$ &Ave& Max&Min & Ave&Max&Min \\
\hline
0&1.006 &1.010 &1.002 &1.003 &1.008 &1.000 \\
0.3&0.890 &0.897 &0.881 &0.927 &0.933 &0.922 \\
0.6&0.788 &0.795 &0.783 &0.850 &0.856 &0.845 \\
0.9&0.661 &0.665 &0.655 &0.741 &0.746 &0.735 \\
\hline
\end{tabular}
\end{center}
}}
\end{table}

\section{Summary}

 In this article, we introduced the novel 
geometrical index $\delta_\agg$, which is
 associated with the Euler number and is obtained by an 
image processing procedure
for a given digital picture of
aggregated particles   
such  that $\delta_\agg$
represents the degree of the agglomerations of the particles.
Following the algorithm in \cite{MSW},
we constructed digital pictures of aggregated particles   
controlled by the agglomeration parameter $\gamma_\agg \in (0,1)$
 as a Monte-Carlo simulation.
By applying the image processing procedure to the pictures,
we showed that $\delta_\agg$ statistically
reproduces $\gamma_\agg$.
Since we have the data of the center points of the particles,
we also computed the well-established Clark-Evans index
 and showed that it also represents $\gamma_\agg$  well.
However though the methods in the point process analysis including
the Clark-Evans index require
the data of the configuration of the points,
the determination of the center points of the particles
for a given picture is basically an ill-posed problem.
Hence our method has an advantage because
we do not need to find the center points in the computation
of $\delta_\agg$.
In other words, our purpose that we recover 
 $\gamma_\agg$ for a given picture by means of
the  digital image processing procedure is accomplished
by considering the deformation of the geometrical object.
It implies that we can measure the 
agglomeration in a given picture of
agglomerated particles.

\bigskip

In this article, we have investigated 
pictures whose  volume fraction $p$ 
is less than $0.5$, because it is difficult to deal with
pictures with large volume fraction.
It is expected to find further natural index to discriminate
the agglomeration with the large volume fraction, e.g., 
in terms of the persistent homology \cite{MPS}.

Further in \cite{KMM},
T. Kaczynski, K. Mischaikow and M. Mrozek
studied the pattern analysis in a regular lattice
using the cubical homology.
They also investigated a topological property of
the time development of a complicated pattern governed by 
the Cahn-Hilliard equation by considering
its time development of its Betti numbers \cite{KMM}.
It means that
a topological property of (geometrical or physical)
deformation of a complicated geometrical
object is important in order to
describe the degree of its complication.
\bigskip

\section*{Acknowledgment}
The authors are grateful to Professors 
Y. Fukumoto and Y. Hiraoka for their
critical and helpful comments, and thank
the referee for critical comments
and for directing their attention to the reference \cite{KMM}.

%\end{document}
\bigskip
\noindent
Shigeki Matsutani, Yoshiyuki Shimosako\\
Simulation \& Analysis R\&D Center,\\
Canon Inc., 3-30-2, Shimomaruko Ohta-ku,\\
Tokyo, Japan
\end{document}